\documentclass[draft]{agujournal2019}
\usepackage{url}
\usepackage{amsmath, braket, subfig, lmodern, float, relsize}
\usepackage[finalnew]{trackchanges} 
\usepackage{soul}
\addeditor{PP}

\draftfalse
\journalname{Journal of Advances in Modeling Earth Systems (JAMES)}

\begin{document}
\title{JAX-LaB: A High-Performance, Differentiable, Lattice Boltzmann Library for Modeling Multiphase Fluid Dynamics in Geosciences and Engineering}


\authors{Piyush Pradhan\affil{1}, Pierre Gentine\affil{1} and Shaina Kelly\affil{1}}

\affiliation{1}{Department of Earth and Environmental Engineering, Columbia University, New York, NY, USA}

\correspondingauthor{Pierre Gentine}{pg2328@columbia.edu}
\correspondingauthor{Shaina Kelly}{sak2280@columbia.edu}



\begin{keypoints}
\item We present a JAX-based differentiable multiphase Lattice Boltzmann library, scalable across CPUs, GPUs, and distributed systems.
\item A modified Shan-Chen pseudopotential incorporating arbitrary equations of state and tunable wettability and surface tension is employed.
\item The library features decoupled compute and write precision, and achieves giga-scale lattice updates per second on multi-GPU setups.
\end{keypoints}

%
%

%
%

\begin{abstract}
We introduce JAX-LaB, a differentiable, Python-based Lattice Boltzmann simulation library designed for modeling multiphase and multiphysics fluid dynamics problems in hydrologic, geologic, and engineered porous media settings. The library is designed as an extension to XLB \cite{ataei_xlb_2024}, and it is built on the JAX framework \cite{jax2018github}. The library delivers a performant, hardware-agnostic implementation that seamlessly integrates with machine learning libraries and scales efficiently across CPUs, multi-GPU setups, and distributed environments. Multiphase interactions are modeled using the Shan-Chen pseudopotential method, coupled with an equation of state (EOS) to reproduce densities consistent with Maxwell's construction, enabling accurate simulation of flows with density ratios $> 10^7$ while maintaining low spurious currents. Fluid wetting is achieved using the "improved" virtual density scheme \cite{li_implementation_2019}, which enables precise control of contact angle on flat and curved surfaces, while eliminating non-physical films seen in the Shan-Chen virtual density scheme. This scheme integrates directly into the interaction force calculations, removing the need to handle fluid-fluid and fluid-solid forces separately. We validate the library's accuracy and performance through comprehensive analytical benchmarks, including Laplace's law, capillary rise in parallel plates, and multi-component cocurrent flow in a channel. We then use the code for several applications involving multicomponent and multiphase flows, including permeability estimation, injection of supercritical $\mathrm{CO_2}$ in a water-saturated Fontainebleau sandstone, and obtaining the characteristic curves for a sphere pack geometry. \change[PP]{Finally, the single-GPU performance and multi-GPU scaling of the code are evaluated}{Finally, the single-GPU performance and multi-GPU scaling of the code are evaluated on both single-node and distributed systems.} The library is open-source under the Apache license and available at \url{https://github.com/piyush-ppradhan/JAX-LaB}.
\end{abstract}

\section*{Plain Language Summary}
Several natural and engineering processes can be classified as multiphase flows involving either a single chemical entity (for instance, water) in multiple phases (typically liquid and gas) or distinct chemical entities such as water and air coexisting in distinct phases. For many such applications, numerical simulations are necessary, as experimental investigations can be challenging. To accurately model multiphase flows, it is essential not only to solve the governing equations that describe fluid dynamics, but also to properly capture phase distributions and account for complex fluid behaviors such as liquid-vapor interfaces, which continuously form, deform, and break during flow, and wettability, which governs fluid-solid interactions. In this work, we introduce a Python-based library to model multiphase systems. The library performs computation using JAX, a high-performance Python library for scientific computations. We describe the various numerical techniques implemented in the library and then benchmark them against several well-known problems with analytical solutions, highlighting the applicability of JAX-LaB for modeling many different multiphase fluid flows. We then showcase several exemplary problems where JAX-LaB can be used in the geosciences and beyond. Finally, we test the performance of the library on single and multiple Graphics Processing Units (GPUs).

\section{Introduction}
Natural and engineered fluid‐flow processes frequently involve more than one phase or chemical species interacting simultaneously. For example, in nature, water migrates through soil as both liquid and vapor during evapotranspiration; permafrost ice thaws to release liquid water and gas \cite{song_lattice_2016}; microscopic droplets nucleate and grow into clouds \cite{yoshino_lattice_2022}; and large‐scale ocean currents develop from coupled density and temperature gradients \cite{linhao_wind-driven_2005}. In engineering contexts, multiphase transport underpins enhanced hydrocarbon recovery \cite{xu_pore-scale_2020}, phase‐change desalination membranes separate fresh water from brine \cite{wang_machine_2024}, and atomized fuel droplets evaporate and ignite within combustion chambers \cite{pitsch_large-eddy_2006}. Flows in which a single substance (e.g., water) coexists in multiple phases (for instance, liquid and vapor) are termed multiphase flows, whereas those involving distinct chemical components (e.g., $\mathrm{CO_2}$ and water) in several phases are termed multicomponent multiphase flows. Both exhibit complex behaviors such as surface tension, capillarity, phase change (evaporation and condensation), and, in some cases, cavitation that govern mass, momentum, and energy transport across scales from pore spaces to the atmosphere or industrial reactors \cite{sukop_lattice_2006}. Interfaces between these phases continually migrate, deform, break up, and merge, and the intricate geometries (e.g., porous rocks or soils) and extreme conditions (e.g., supercritical fluids) in many real-world systems often make experimental studies impractical. As a result, numerical modeling has become an essential tool for investigating multiphase flows \cite{aidun_lattice-boltzmann_2010}, provided it accurately captures vapor-liquid fractions and their densities via an appropriate equation of state and reproduces interface dynamics and wetting behavior of fluids.

Python libraries such as NumPy \cite{harris_array_2020}, JAX \cite{jax2018github}, PyTorch \cite{paszke_pytorch_2019} and TensorFlow \cite{abadi_tensorflow_2016} provide a high-level interface for performing computations on arrays (tensors) while leveraging highly optimized C/C++ libraries in the backend to target massively parallel hardware such as graphics processing unit (GPU) and tensor processing unit (TPU). This approach combines the performance of C/C++ with the ease of using Python and hence, finds extensive use in domains such as scientific computing and machine learning. Recent works from \citeA{kochkov_machine_2021, vinuesa_enhancing_2022} have highlighted the potential of machine learning to accelerate the performance of Computational Fluid Dynamics (CFD) codes and in predicting the spatio-temporal evolution of fluid physics, and this development is motivated by the emergence of powerful general-purpose automatic differentiation frameworks such as Tensorflow \cite{abadi_tensorflow_2016}, JAX \cite{jax2018github}, and PyTorch \cite{paszke_pytorch_2019}. Yet, to date, numerical models for multicomponent multiphase flows have not been well integrated with ML techniques, which is primarily due to the use of different programming languages for machine Learning (Python) and CFD (C/C++/Fortran). Thus, there is a need to develop a single-language implementation of a high-performance, accurate, and differentiable CFD code that can integrate well with existing machine learning libraries to provide a unified workflow for forward and inverse modeling in CFD.

The Lattice Boltzmann Method (LBM) is a widely used computational approach for simulating fluid flows across diverse scales and applications, including porous media, microfluidics, biological flows, and even large/exascale applications such as meteorological flows \cite{mittal_immersed_2005, zhang_lattice_2011, worner_numerical_2012, aidun_lattice-boltzmann_2010, petersen_lattice_2021, feng_hybrid_2019}. LBM distinguishes itself from other well-known CFD methods, such as the finite volume method (FVM), by numerically solving a discretized form of the Boltzmann equation, known as the Lattice Boltzmann Equation (LBE), instead of the Navier-Stokes equation. The main advantage of the Lattice Boltzmann Method (LBM) is its simple two-step algorithm: a local collision step followed by a streaming step that transfers data to neighboring nodes. This structure, along with its explicit time integration, makes LBM inherently parallel and highly efficient on modern accelerated hardware such as Graphics Processing Units (GPUs). Additionally, LBM utilizes uniform Cartesian grids instead of geometry-conforming meshes, making it an ideal choice for modeling complex or heterogeneous domains. Using the Shan-Chen pseudopotential method, LBM allows easy incorporation of numerous features to model real-world fluids, such as equation of state \cite{yuan_equations_2006}, surface tension control \cite{hu_surface_2014, li_achieving_2013} and interface dynamics such as formation, merging and deformation without performing computationally-expensive interface tracking \cite{coon_taxila_2014}. Moreover, LBM can be used to simulate turbulent flows, either directly (DNS) using specific collision model, such as the entropic KBC model \cite{karlin_gibbs_2014} or by incorporating turbulence models such as Reynolds-averaged Navier Stokes (RANS) \cite{filippova_multiscale_2001, li_non-body-fitted_2012, pellerin_implementation_2015} and large eddy simulation (LES) \cite{premnath_dynamic_2009, bartlett_lattice_2013}. In the context of hydrology, LBM has been effectively used to model evaporation from porous media \cite{fei_pore-scale_2022, tian_lattice_2025}, thawing of frozen soils \cite{song_lattice_2016}, sediment transport in soils \cite{zhou_pore-scale_2021}, cloud formation \cite{yoshino_lattice_2022} and ocean circulation \cite{linhao_wind-driven_2005}. In the context of geohydrology or subsurface reservoir engineering, LBM has been effectively used to understand immiscible displacement, drainage, imbibition, and residual trapping during enhanced oil recovery from reservoirs \cite{liu_review_2021}, flow of supercritical carbon dioxide ($\mathrm{CO_2}$) through rocks \cite{xie_relative_2017, an_lattice-boltzmann_2021}, identification of the representative elementary volume (REV) in low-permeability reservoirs such as shales \cite{kelly_assessing_2016}, and to validate continuum-scale models by deriving effective properties like permeability, capillary pressure functions and relative permeability curves \cite{liu_review_2021, zhang_flow_2019}. LBM can be coupled with other numerical solvers to model multiphysics systems. For example, LBM can be extended to handle the energy equation (thermal LBM) for simulating evaporation and boiling phenomena \cite{luo_unified_2021}, or combined with the advection-diffusion equation to model reactive transport in porous media \cite{gao_reactive_2017}. Moreover, LBM is well-suited for array-based computation, making it highly compatible with machine learning frameworks for both forward and inverse modeling tasks. These include applications such as permeability estimation \cite{graczyk_predicting_2020}, steady-state flow prediction \cite{ataei_xlb_2024, wang_ml-lbm_2021}, and capturing hysteresis behavior in porous materials \cite{chaaban_machine-learning_2024}. Machine learning can also enhance the performance of LBM by optimizing the relaxation rates according to the spatio-temporal characteristics of the flow field \cite{bedrunka_machine_2024, horstmann_lattice_2024}. Finally, fast LBM solvers enable large-scale data generation, which can serve as training datasets for data-driven models.

Open-source libraries such as Palabos \cite{latt_palabos_2021}, OpenLB \cite{krause_openlbopen_2021}, MF-LBM\cite{chen_inertial_2019}, Taxila-LBM \cite{coon_taxila_2014}, Sailfish \cite{januszewski_sailfish_2014}, waLBerla \cite{bauer_walberla_2021}, MPLBM-UT \cite{santos_mplbm-ut_2022} and ch4-project \cite{calzavarini_eulerianlagrangian_2019} can run LBM simulation in parallel across CPUs and GPUs, but are written using C/C++ or Fortran and rely on vendor-specific libraries such as CUDA \cite{luebke_cuda_2008}. Platform-agnostic codes such as FluidX3D \cite{lehmann_high_2019}, which is based on the Open Compute Language (OpenCL) \cite{stone_opencl_2010}, also exist; however, it too is written in C++. In contrast, JAX-LaB, built as an extension of the single-phase differentiable LBM library XLB \cite{ataei_xlb_2023}, offers a Python-based solution that seamlessly targets various backends like CPU, GPU, multi-GPU, and distributed GPU systems without vendor-specific code modifications, enabling rapid prototyping and execution while allowing for easy solver customization. Furthermore, in contrast to other Python-based differentiable CFD libraries such as Lettuce \cite{bedrunka_lettuce_2021}, Taichi-LBM3D \cite{yang_taichi-lbm3d_2022}, and JAX-Fluids (which is based on Finite Volume Method) \cite{bezgin_jax-fluids_2023, bezgin_jax-fluids_2025}, JAX-LaB supports both single and multicomponent multiphase systems by incorporating any non-ideal equation of state and delivers superior performance on single-GPU platforms compared to Lettuce and Taichi-LBM3D.

\section{Numerical Methods and Implementation}
JAX-LaB models fluids using the lattice Boltzmann method (LBM), which is paired with the Shan-Chen pseudopotential method to model multiphase-multicomponent systems. The pseudopotential computation is modified from the original formulation by \citeA{shan_lattice_1993} to incorporate non-ideal fluid behavior using an equation of state(EOS), and control the interfacial tension and fluid-solid wettabilty for multiple fluid phases. The code implements commonly used equations of state, including Carnahan-Starling, Peng-Robinson, Redlich-Kwong, Redlich-Kwong-Soave, and Van der Waals, enabling the modeling of high-density ratio flows ($> 10^8$) with minimal spurious currents ($< 2 \times 10^{-3}$) for a wide range of contact angles ($5-170^\circ$). In this section, the numerical methodology employed in JAX-LaB, including the available collision models and multiphase implementation, is described.


\subsection{Lattice Boltzmann method}
Lattice Boltzmann method (LBM) describes the evolution of the probability distribution function $f$ by solving the lattice Boltzmann equation (LBE):
\begin{equation}\label{eq:lbe}
    \begin{split}
        f_i(x + \mathbf{e_i} \Delta t, t + \Delta t) - f_i(x, t) &= \Omega_i \hspace{0.2cm} i = 0, 1 ... q-1
    \end{split}
\end{equation}
Here, $\Omega$ is the collision operator. Distributions are evaluated at a fixed set of points on a Cartesian grid and for specific velocity directions ($\mathbf{e}_i$), typically defined using the D\textit{d}Q\textit{q} scheme, where \textit{d} represents the dimension and \textit{q} denotes the number of velocity directions considered. Some popular lattice velocity sets used in LBM include D2Q5 \& D2Q9 in two dimensions and D3Q15, D3Q19 \& D3Q27 in three dimensions. The LBE is numerically solved using a combination of \textit{collision} (local) and \textit{streaming} (non-local) steps. In the \textit{collision} step, the distribution functions for the next timestep are determined using a collision model, while in the \textit{streaming} step, these distributions are moved to their respective adjacent nodes along the directions defined in the discrete velocity set.
\begin{equation}\label{eq:lbe_steps}
    \begin{split}
        &\mathrm{collision}: f_i^{*}(\mathbf{x}, t) = f_i + \Omega_i \\
        &\mathrm{streaming}: f_i(\mathbf{x} + \mathbf{e_i} \Delta t, t + \Delta t) = f_i^{*}(\mathbf{x}, t)
    \end{split}
\end{equation}
JAX-LaB supports various collision operators such as the single-relaxation time or Bhatnagar-Gross-Krook (BGK) operator, multi-relaxation time (MRT) operator \cite{coveney_multiplerelaxationtime_2002}, and the central moment (cascaded) operator (CLBM) \cite{geier_cascaded_2006}. The post-collision distribution using these collision models can be summarized as:
\begin{equation}\label{eq:collision_model}
\begin{split}
    &\mathrm{BGK}: f_i^{*}(\mathbf{x}, t) - f_i(\mathbf{x}, t) = -\frac{1}{\tau} [f_i(\mathbf{x}, t) - f_i^{eq}(\mathbf{x}, t)] \\
    &\mathrm{MRT}: f_i^{*}(\mathbf{x}, t) - f_i(\mathbf{x}, t) = -(\mathbf{M}^{-1} \mathbf{S} \mathbf{M}) [f_i(\mathbf{x}, t) - f_i^{eq}(\mathbf{x}, t)] \\
    &\mathrm{CLBM}: f_i^{*}(\mathbf{x}, t) - f_i(\mathbf{x}, t) = -(\mathbf{M}^{-1} \mathbf{N}^{-1} \mathbf{S} \mathbf{N} \mathbf{M}) [f_i(\mathbf{x}, t) - f_i^{eq}(\mathbf{x}, t)] \\
\end{split}
\end{equation}
The macroscopic variables such as density $\rho$ and velocity $\mathbf{u}$ are computed using the moments of $f$:
\begin{equation}\label{eq:macroeq}
    \begin{split}
        \rho & = \sum_{i = 0}^{q-1} f_i \\
        \rho \mathbf{u} &= \sum_{i = 0}^{q-1} f_i \mathbf{e_i}
    \end{split}
\end{equation}
In JAX-LaB, for BGK and MRT collision models, the external forces are incorporated into the distribution function using the Exact Difference Method (EDM) \cite{kupershtokh_equations_2009}, which modifies the post-collision distribution function using a difference of equilibrium distribution functions computed using the momentum before and after the action of force $\mathbf{F}$:
\begin{equation}\label{eq:edm}
    \begin{split}
        \Delta \mathbf{u} &= \frac{\mathbf{F} \Delta t}{\rho} \\
        \mathbf{F_m} &= f_i^{eq}(\rho, \mathbf{u} + \Delta\mathbf{u}) - f_i^{eq}(\rho, \mathbf{u}) \\
        f_i^{*}(\mathbf{x}, t) &= f_i^{*}(\mathbf{x}, t) + \mathbf{F_m}
    \end{split}
\end{equation}
In this case, the macroscopic velocity is computed as:
\begin{equation}
    \begin{split}
        \rho\mathbf{u} = \sum_i f_i \mathbf{e}_i + \frac{\mathbf{F}\Delta t}{2}
    \end{split}
\end{equation}
Although EDM is compatible with any collision operator as it is directly derived from the Boltzmann equation \cite{kupershtokh_equations_2009}, for the cascaded collision model, a consistent forcing scheme based on second-order trapezoidal scheme proposed by \citeA{fei_consistent_2017} is used for enhanced performance, which introduces an additional forcing term to the central moments:
\begin{equation}
\begin{gathered}
        f_i^{*}(x, t) - f_i(x, t) = \mathbf{M}^{-1} \mathbf{N}^{-1} (-\mathbf{S} \mathbf{N} \mathbf{M} [f(\mathbf{x}, t) - f_i^{eq}(\mathbf{x}, t)] +(\mathbf{I} - \mathbf{S}/2)\mathbf{C}) \\
    \mathbf{C} = \mathbf{N}\mathbf{M}\mathbf{R}_i \\
    \mathbf{R}_i = \frac{\mathbf{F}}{\rho}\frac{(\mathbf{e}_i - \mathbf{u})}{c_s^2} f_i^{eq}
\end{gathered}
\end{equation}
Similar to EDM, the macroscopic velocity is computed as:
\begin{equation}
    \begin{split}
        \rho\mathbf{u} = \sum_i f_i \mathbf{e}_i + \frac{\mathbf{F}\Delta t}{2}
    \end{split}
\end{equation}
\subsection{Multiphase \& Multicomponent Flows}\label{multiphase-implementation}
For multicomponent flows, the evolution of distribution $f$ for the $k$-th component is obtained using LBE:
\begin{equation}\label{eq:mcmp_lbe}
    \begin{split}
        f_i^k(x + \mathbf{e_i} \Delta t, t + \Delta t) - f_i^k(x, t) &= \Omega_i^k \hspace{0.2cm} i = 0, 1 ... q-1
    \end{split}
\end{equation}
The corresponding fluid density and velocity are obtained using Eq. \ref{eq:macroeq}:
\begin{equation}\label{eq:macroeq_mcp}
    \begin{split}
        \rho^k(\mathbf{x}, t) &= \sum_i f_{i}^{k} (\mathbf{x}, t) \\
        \mathbf{u}^k(\mathbf{x}, t) &= \sum_i f_{i}^{k} (\mathbf{x}, t) \mathbf{e}_i / \rho^k(\mathbf{x}, t)
    \end{split}
\end{equation}
The interactions among phases and components are modeled using the Shan-Chen pseudopotential method \cite{shan_lattice_1993}, chosen for several key advantages: (i) it is simple and computationally efficient, relying on a localized interaction force to induce phase separation, eliminating the need for explicit interface tracking \cite{gong_numerical_2012, coon_taxila_2014} or evolution of a separate order-parameter field (which is used in the free-energy method \cite{swift_lattice_1996}); (ii) it naturally supports both single-component and multicomponent multiphase systems unlike the color-gradient method \cite{gunstensen_lattice_1991} which is limited to binary fluid mixtures; (iii) it accommodates a broader range of density ratios than the color-gradient approach; (iv) it preserves Galilean invariance, which the free-energy method does not; and (v) it allows for the use of arbitrary equations of state, whereas the free-energy method has primarily been demonstrated with the van der Waals EOS \cite{gong_numerical_2012}. Despite these strengths, the Shan-Chen method has some limitations. It typically produces diffuse interfaces spanning several voxels, in contrast to the sharp, well-defined interfaces generated by the color-gradient method \cite{chen_inertial_2019}. Additionally, the Shan-Chen approach requires tuning the pressure modification parameter $\kappa$ (Section \ref{surface_tension_adjust}) to indirectly control interfacial tension, while surface tension can be directly specified in the color-gradient model.

The force $F_{f-f}^{k,k'}$ represents the interaction between $k$ and $k'$th fluid components and it is modeled as:
\begin{equation}\label{eq:scmp_fluid_fluid_force}
    \begin{gathered}
        F_{f-f}^{k, k'}(\mathbf{x}, t ) = -A_{kk'} \nabla U^{k^{'}}(\mathbf{x}, t) - (1 - A_{kk'}) g_{kk'} \psi^k(\mathbf{x}, t) \nabla \psi^{k^{'}}(\mathbf{x}, t)
    \end{gathered}
\end{equation}
Eq. \ref{eq:scmp_fluid_fluid_force} is the weighted sum of forces obtained from two different forcing schemes: the original Shan-Chen forcing scheme \cite{shan_lattice_1993} and the Zhang-Chen forcing scheme \cite{zhang_lattice_2003}. This generalized approximation of the interaction force is widely used in the literature as it can model systems with a broader range of density ratios compared to each of the individual schemes \cite{hu_surface_2014,qin_effective_2022, gong_numerical_2012, kupershtokh_equations_2009}.

To model the behavior of non-ideal fluids, pressures obtained from a given equation of state (EOS) are used to compute the pseudopotentials used in Eq. \ref{eq:scmp_fluid_fluid_force} \cite{yuan_equations_2006, hu_surface_2014, kupershtokh_equations_2009}:
\begin{equation}\label{eq:eos_potential}
    \begin{gathered}
        \psi^k(\rho) = \sqrt{\frac{2(\alpha^k p^k_{EOS} - c_s^2 \rho^k)}{g_{kk}}} \\
        U^k(\mathbf{x}, t) = \alpha^k p^k_{EOS}(\mathbf{x}, t) - c_s^2 \rho^k(\mathbf{x}, t)
    \end{gathered}
\end{equation}
These equations of state are implemented under a separate EOS class in code (\verb|src/eos.py|), which can be appended by the user to add other equations of states (for instance, cross-over EOS to model supercritical fluids \cite{ashirbekov_equation_2021}). To run a multiphase simulation, an EOS object is passed as an argument to the consolidated multiphase parent class, and it is subsequently used by the pseudopotential computation function. This approach provides a convenient and elegant solution to incorporate an arbitrary EOS, as the EOS object can be easily replaced by another to use a different equation of state without making any modification to the multiphase code. The interaction strength $g_{kk'}$ controls the miscibility between different fluids. The intra-component strength $g_{kk}$ is set as -1 to ensure that the term inside the square root stays positive \cite{yuan_equations_2006}. The total force for the $k$th component is obtained by summing over all $F^{k, k'}$. The gradients of pseudopotentials $\psi$ and $U$ are approximated using a weighted sum of their values at neighboring lattice sites:
\begin{equation}\label{eq:software_ff_force_mcmp}
    \begin{split}
        \mathbf{F}_{f-f}^k &= A_{kk} \mathbf{F}_{ZC}^k + (1 - A_{kk}) \mathbf{F}_{SC}^k \\
        \mathbf{F}_{SC}^k &=  -\psi^k(\mathbf{x}) \sum_{k'=1}^{n_C} g_{kk'} \sum_{\mathbf{x'}} G(\mathbf{x}, \mathbf{x'}) \psi^{k'}(\mathbf{x'}) \mathbf{(x' - x)} \\
        \mathbf{F}_{ZC}^k &= -\sum_{\mathbf{x'}} G(\mathbf{x}, \mathbf{x'}) U^k(\mathbf{x'}) \mathbf{(x' - x)}
\end{split}
\end{equation}
By default, the Green's function $G(\mathbf{x}, \mathbf{x'})$ specifies non-zero values only for nearest neighbor sites. However, users can specify non-zero values for next-nearest neighbors to achieve higher-order isotropy for the gradient approximation \cite{coon_taxila_2014, qin_effective_2022, sbragaglia_generalized_2007}. 
\subsection{Interfacial tension control}\label{surface_tension_adjust}
In the Shan-Chen pseudopotential model, surface tension and density ratio are intrinsically linked, making it impossible to adjust one without affecting the other \cite{shan_lattice_1993}. Although adjusting the modification parameter $\alpha^k$ and the weighting coefficients $A_{kk'}$ allows for some control over surface tension, the tunable range is limited (typically 2-3 times of unadjusted value) without compromising the ability to reproduce the correct coexistence densities \cite{hu_surface_2014}. To overcome this constraint, JAX-LaB implements the pressure tensor modification approach developed by Li et al. \cite{li_achieving_2013} to tune the surface tension without changing the mechanical stability of the multiphase system. In this method, the MRT collision step described in Eq. \ref{eq:collision_model} is modified to incorporate a source term $\mathbf{C}$:
\begin{equation}\label{eq:mrt_surface}
\begin{split}
    f_i^{k}(x + \mathbf{e}_i \Delta t, t + \Delta t) - f_i^{k}(x + \mathbf{e}_i \Delta t, t) &= -\mathbf{M}^{-1}[(\mathbf{S}\mathbf{M}) (f_i^{k} - f_i^{k, eq}) - F_{m} ] + \mathbf{C}
\end{split}
\end{equation}
For D2Q9 lattice, $\mathbf{C}$ is given as \cite{li_achieving_2013}:
\begin{equation}\label{eq:li_surface_adjust}
\begin{split}
    \mathbf{C} &= [0, 1.5s_e(Q_{xx} + Q_{yy}), -1.5s_{\eta}(Q_{xx} + Q_{yy}), 0, 0, 0, 0, -s_{\nu}(Q_{xx} - Q_{yy}), -s_\nu Q_{xy}]^{T}
\end{split}
\end{equation}
where the terms $Q_{xx}$, $Q_{yy}$ and $Q_{xy}$ are evaluated using \cite{hu_surface_2014}:
\begin{equation}
\begin{split}
    \mathbf{Q} & = \kappa[(1-A_{kk})\psi(\mathbf{x})\sum_{i=0}^{q-1} w(|\mathbf{e}_\alpha|^2)(\psi^{k}(\mathbf{x} + \mathbf{e}_\alpha) - \psi^{k}(\mathbf{x}))\mathbf{e}_\alpha\mathbf{e}_\alpha + \\ &\frac{A_{kk}}{2}\sum_{i=0}^{q-1} w(|\mathbf{e}_\alpha|^2)(\psi^{k}(\mathbf{x} + \mathbf{e}_\alpha)^2 - \psi^{k}(\mathbf{x})^2)\mathbf{e}_\alpha\mathbf{e}_\alpha]
\end{split}
\end{equation}
$\mathbf{Q}$ is based on the discrete pressure tensor formulation for the pseudopotential method \cite{shan_pressure_2008}. The modified pressure tensor obtained after the addition of the source term causes the surface tension to scale with $1 - \kappa$, reaching its maximum magnitude when this term equals unity. The formulation of source $\mathbf{C}$ for D3Q19 lattice can be found in \citeA{zhu_investigation_2024}.
\subsection{Wettability control}
In JAX-LaB, solid-fluid wettability is implemented using the improved virtual density scheme proposed by \citeA{li_implementation_2019}. This method simulates wettability by locally modifying density values at solid nodes:
\begin{equation}\label{eq:improved_virtual_density}
\rho_s = \left\{
	\begin{array}{lll}
		\phi \rho_\mathrm{ave}(x_s), & \phi \geq 1, & \theta \leq 90^\circ \\
		\rho_\mathrm{ave}(x_s) - \Delta \rho, & \Delta \rho \geq 0,  & \theta > 90^\circ
	\end{array}
\right.
\end{equation}
where $\rho_s$ is the density at solid nodes $x_s$, $\phi$ \& $\Delta \rho$ are control parameters and $\rho_{ave}$ is a weighted average of neighboring fluid densities \cite{li_implementation_2019}. To ensure numerical stability, a limiter constrains the solid node densities from Eq. \ref{eq:improved_virtual_density} to lie within the problem's physical density bounds. By employing spatially varying solid densities rather than a uniform global value, this scheme retains the strengths of the geometric contact angle method \cite{ding_wetting_2007} without the need to compute characteristic directions, which can be computationally expensive, especially in three dimensions \cite{li_implementation_2019}. In doing so, it overcomes major drawbacks of traditional Shan-Chen wettability mod`els, namely, large spurious currents at high density ratios, limited control over contact angles at high density ratios, and the emergence of non-physical interfacial films whose thickness is tied to density ratio rather than wettability \cite{hu_contact_2016, li_implementation_2019}, while still achieving a broad spectrum of contact angles by tuning $\phi$, $\rho_{\mathrm{ave}}$, and $\Delta \rho$.
\subsection{Model Implementation}
JAX-LaB leverages JAX \cite{jax2018github}, a high-performance hardware-agnostic library from Google that breaks down functions into efficient primitives and uses them to generate machine code for accelerators like GPUs and Tensor Processing Units (TPUs). The functions are traced with abstract arguments (encoding shape and type) on the first call, and the optimized code generated is reused directly for subsequent calls with matching inputs, enabling significant performance gains \cite{jax2018github}. As LBM computations rely on fixed-shape, fixed-precision arrays, JAX-LaB leverages an optimized implementation of the LB step (Eq. \ref{eq:lbe_steps}) for significant performance enhancement.
\subsection{Data Structure Design}\label{mcmp_data_struct}
In XLB \cite{ataei_xlb_2024}, GPU-accessible arrays are defined as JAX arrays with dimensions (x, y, cardinality) in 2D or (x, y, z, cardinality) in 3D. These arrays are distributed across multiple GPUs using a computational sharding strategy that slices the array along the x-axis and assigns each slice to a GPU. To balance the load between different GPUs, \verb|nx| is adjusted to be divisible by the number of GPUs. Furthermore, initialization is restricted to the portion of the array accessible by each GPU, which is crucial for arrays that are too large to be initialized on a single GPU.

\begin{figure}[ht]
    \centering
    \noindent\includegraphics[scale=0.02]{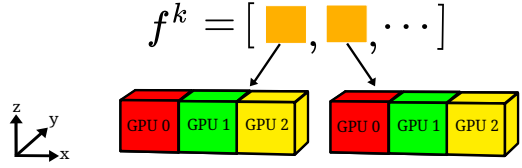}
    \caption{Illustration of computational sharding strategy and data structure employed to store JAX arrays and distribute them equally across multiple devices. Arrays are sliced along the x-axis and then divided across multiple devices. These sharded arrays are stored together in a list to form a pytree, with each JAX array storing values for one component.}
    \label{fig:pytree_data_struct}
\end{figure}

\begin{figure}[ht]
    \centering
    \noindent\includegraphics[scale=0.078]{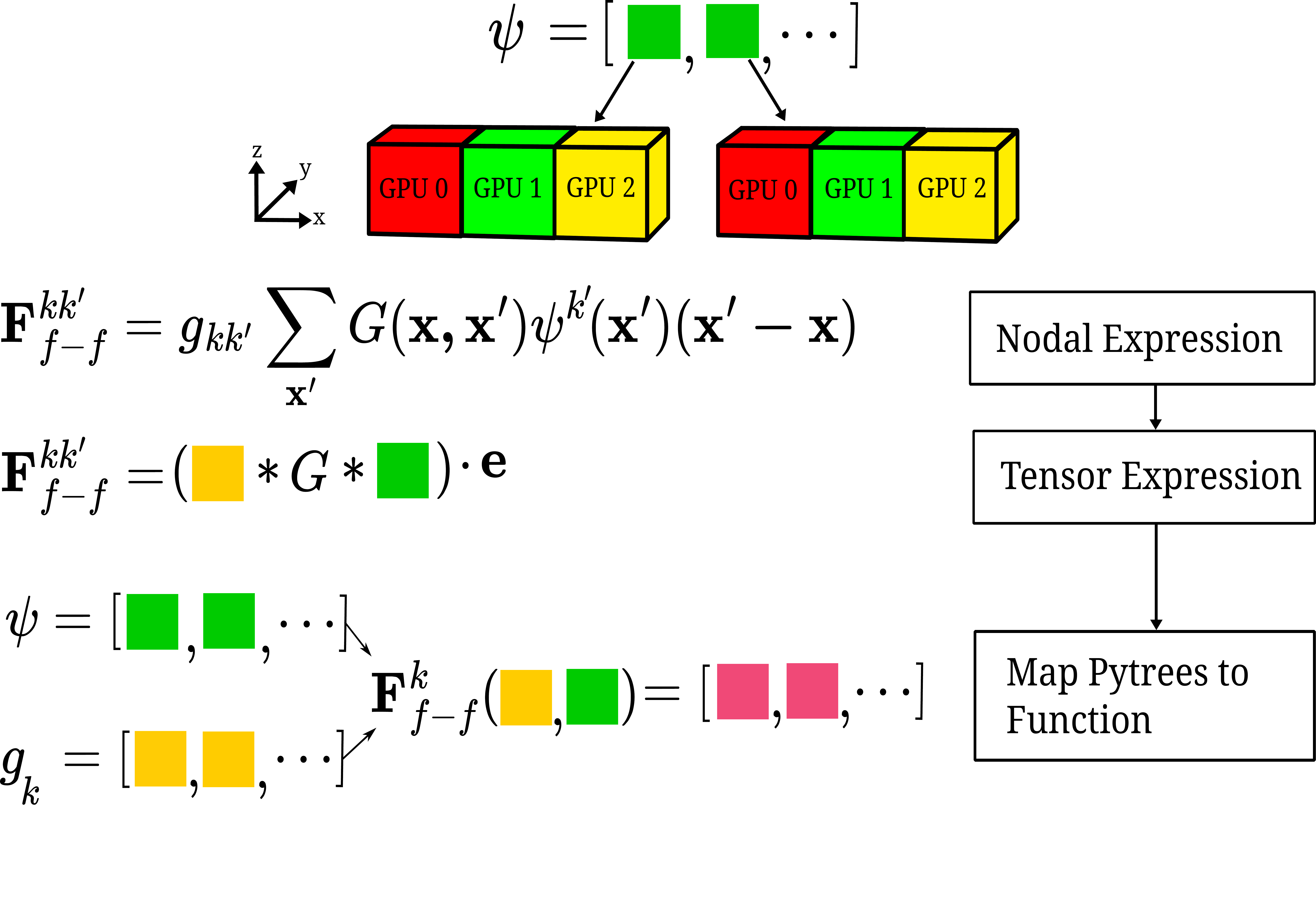}
    \caption{Implementation of Shan-Chen force using pytrees in JAX-LaB. The colored squares indicate the different values of pytree, which can be either a floating-point value or a JAX array.}
    \label{fig:pytree_example}
\end{figure}

JAX-LaB extends these concepts to multicomponent flows by using \textit{pytrees}, a list-based data structure introduced by JAX to store all computational parameters. During initialization, a JAX array is created to store distribution values for each component (using the previously described computational sharding strategy) and collected into a list to form a pytree (Fig. \ref{fig:pytree_data_struct}). To carry out computations, the nodal formulations of various quantities, such as the interaction force, pressure tensor modification, and others, are expressed as equivalent tensor operations. These operations are then applied to the pytrees, producing a new pytree of JAX arrays for all components. This approach is demonstrated using the original Shan-Chen force in Fig. \ref{fig:pytree_example}, and it is used for all computations, leading to a concise formulation that does not change with the number of components. The data structure is also used for storing various floating point values (for instance, fluid-fluid interaction strengths $g_k$ in Fig. \ref{fig:pytree_example}) as it can be passed to functions as a pytree to perform computation.
\section{Benchmarks \& Applications}
To validate JAX-LaB's suitability for modeling real‑world multiphase systems, we tested JAX-LaB against several established analytical benchmarks. These benchmarks examine its ability to (i) reproduce thermodynamically consistent densities per the equation of state (\ref{thermodynamic_const}), (ii) adjust fluid surface tension without altering the system's density ratio (\ref{laplace_law}), and (iii) incorporate wettability in both single‑component and multicomponent multiphase configurations (\ref{washburn}, \ref{channel_drainage}, \ref{cocurrent}). The maximum achievable density ratios for each EOS are dependent on the EOS parameters considered, and in the subsequent section, the performance is evaluated for one set of parameters, which are taken from \citeA{yuan_equations_2006}'s work. The \verb|src/examples| directory in the repository contains scripts for reproducing the benchmarks and applications discussed in this section. The examples presented in Section \ref{applications} highlight the versatility of JAX-LaB for modeling porous media flows.
\subsection{Thermodynamic Consistency}\label{thermodynamic_const}
\begin{figure}[ht]
    \centering
    \noindent\includegraphics[scale=0.48]{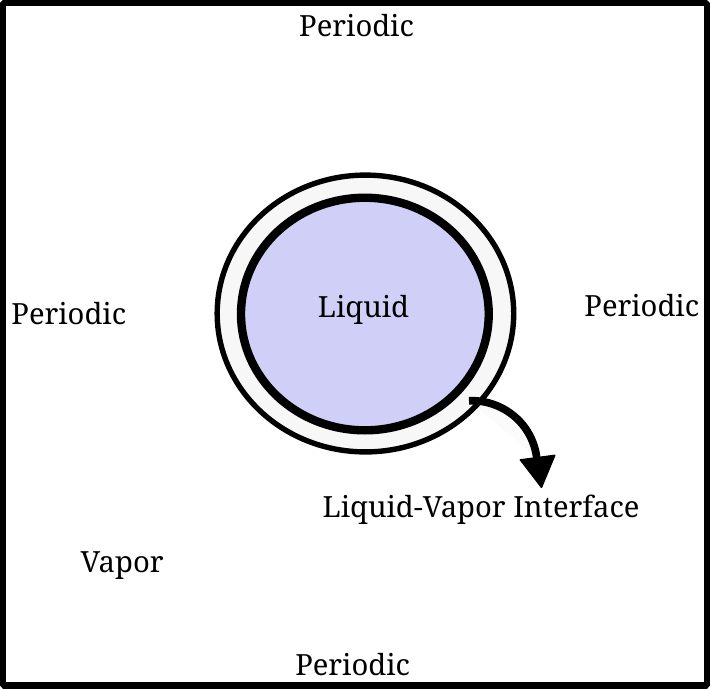}
    \caption{Schematic diagram for a droplet suspended in vapor.}
    \label{fig:droplet_test}
\end{figure}

\begin{figure}[ht]    
\centering
\noindent\includegraphics[scale=0.4]{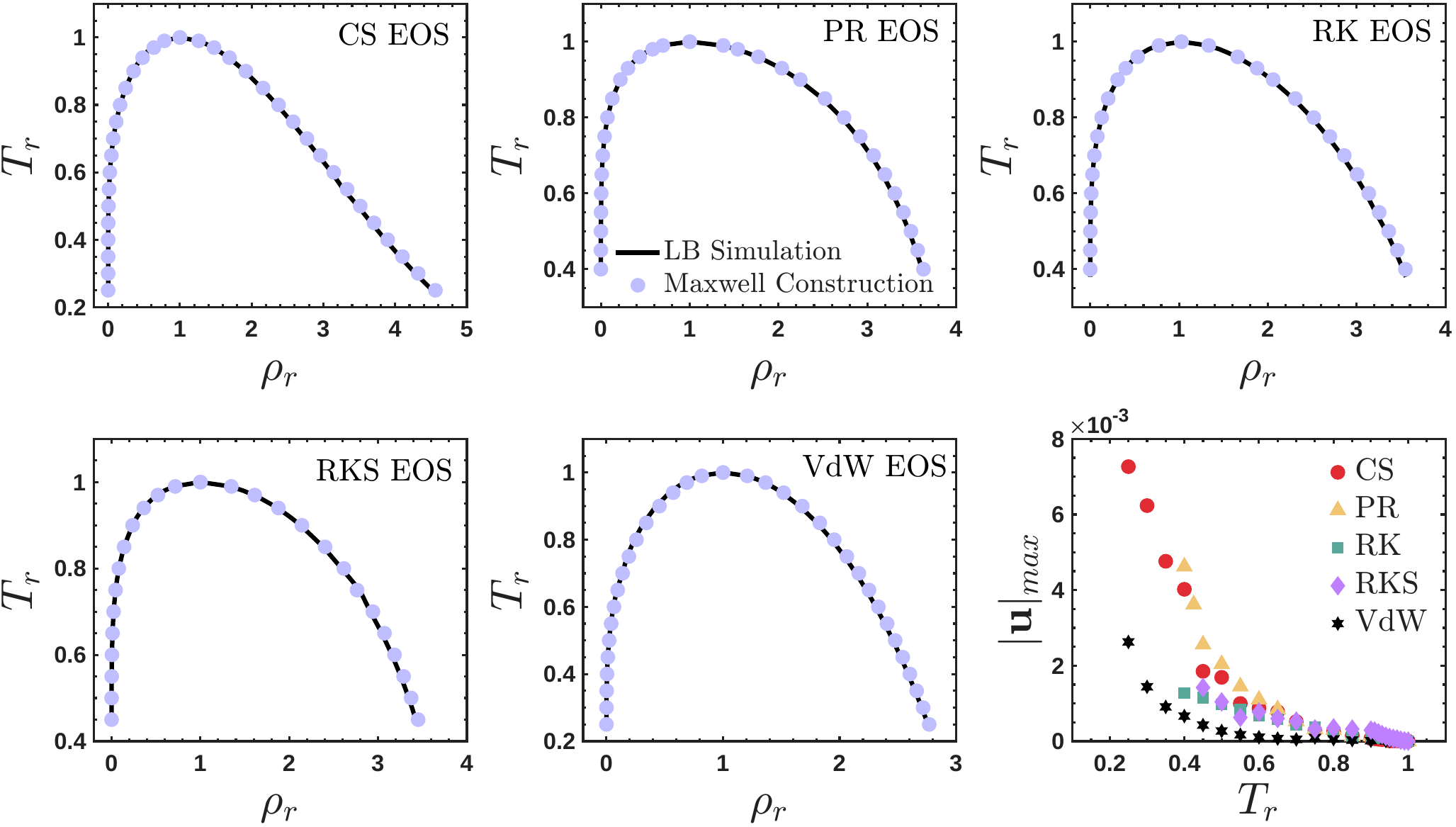}
    \caption{Comparison of coexistence densities versus reduced temperature derived from Lattice Boltzmann simulations with predictions from the Maxwell construction for various equations of state (EOS): (a) Carnahan-Starling (CS) (b) Peng-Robinson (PR) (c) Redlich-Kwong (RK), (d) Redlich-Kwong-Soave (RKS) \& (e) van der Waals (VdW). (f) Peak spurious current magnitudes across various temperatures for different equations of state. All values are in lattice units.}
    \label{fig:eos_laplace_coexistence}
\end{figure}

Coexistence densities are the densities of different phases, such as liquid and vapor, in equilibrium at a specific temperature and pressure. 2D and 3D simulations of a static liquid droplet suspended in its vapor (Fig. \ref{fig:droplet_test}) were performed in JAX-LaB to calculate coexistence densities and verify thermodynamic consistency. In these tests, the parameters for each of the equations of state are taken from \citeA{yuan_equations_2006}. The 2D and 3D simulations were domains of sizes $200^2$ and $200^3$ lattice units, respectively, with a droplet of radius 30 lattice units placed at the domain centers. The initial density profile of the domain was specified as \cite{qin_effective_2022}:
\begin{equation}\label{eq:laplace_init_density}
    \begin{split}
        \rho = \frac{\rho_l - \rho_g}{2} - \frac{\rho_l - \rho_g}{2} \cdot \text{tanh}(2 * \frac{(d -r)}{w})
    \end{split}
\end{equation}
where $d$ is the Euclidean distance of all points in the domain from the droplet center and $w$ is the width of the interface, set as 5 lattice units in the present simulation. The simulations were performed using D2Q9 or D3Q19 lattice with MRT collision model \cite{mccracken_multiple-relaxation-time_2005} and run until steady state was achieved. The values of $\alpha^k$ and $A_kk'$ in Eq. \ref{eq:software_ff_force_mcmp} were adjusted for each EOS to ensure that the mechanical stability condition was satisfied. The coexistence densities obtained from these lattice Boltzmann simulations are compared against the values obtained from Maxwell's construction. In the simulation, five equations of state are considered: Carnahan-Starling (CS), Peng-Robinson (PR), Redlich-Kwong (RK), Redlich-Kwong-Soave (RKS), and Van der Waals (VdW). The coexistence densities (scaled by critical density $\rho_c$ to get reduced densities $\rho_r$) for these EOS are evaluated at different reduced temperatures Tr, and the results are shown in Fig. \ref{fig:eos_laplace_coexistence}. The parameters used for each equation of state are taken from the works of Yuan et.al \cite{yuan_equations_2006}. The Fig. \ref{fig:eos_laplace_coexistence} clearly demonstrates that the coexistence densities obtained from simulation closely match the expected values for a large range of reduced temperature $T_r$, achieving maximum density ratios of $8.439 \times 10^7$, $4.84 \times 10^5$, $2.74 \times 10^5$, $6.77 \times 10^4$ and $5.4 \times 10^4$ for CS, PR, RK, RKS, and VdW equation of states respectively with root mean square (RMS) errors of $2.6 \times 10^{-3}$, $4.2 \times 10^{-2}$, $3.9 \times 10^{-2}$, $4 \times 10^{-2}$ and $5.1 \times 10^{-2}$ respectively.
\subsection{Laplace Law}\label{laplace_law}
\begin{figure}[!ht]
    \centering
    \noindent\includegraphics[width=14cm,height=8.7cm]{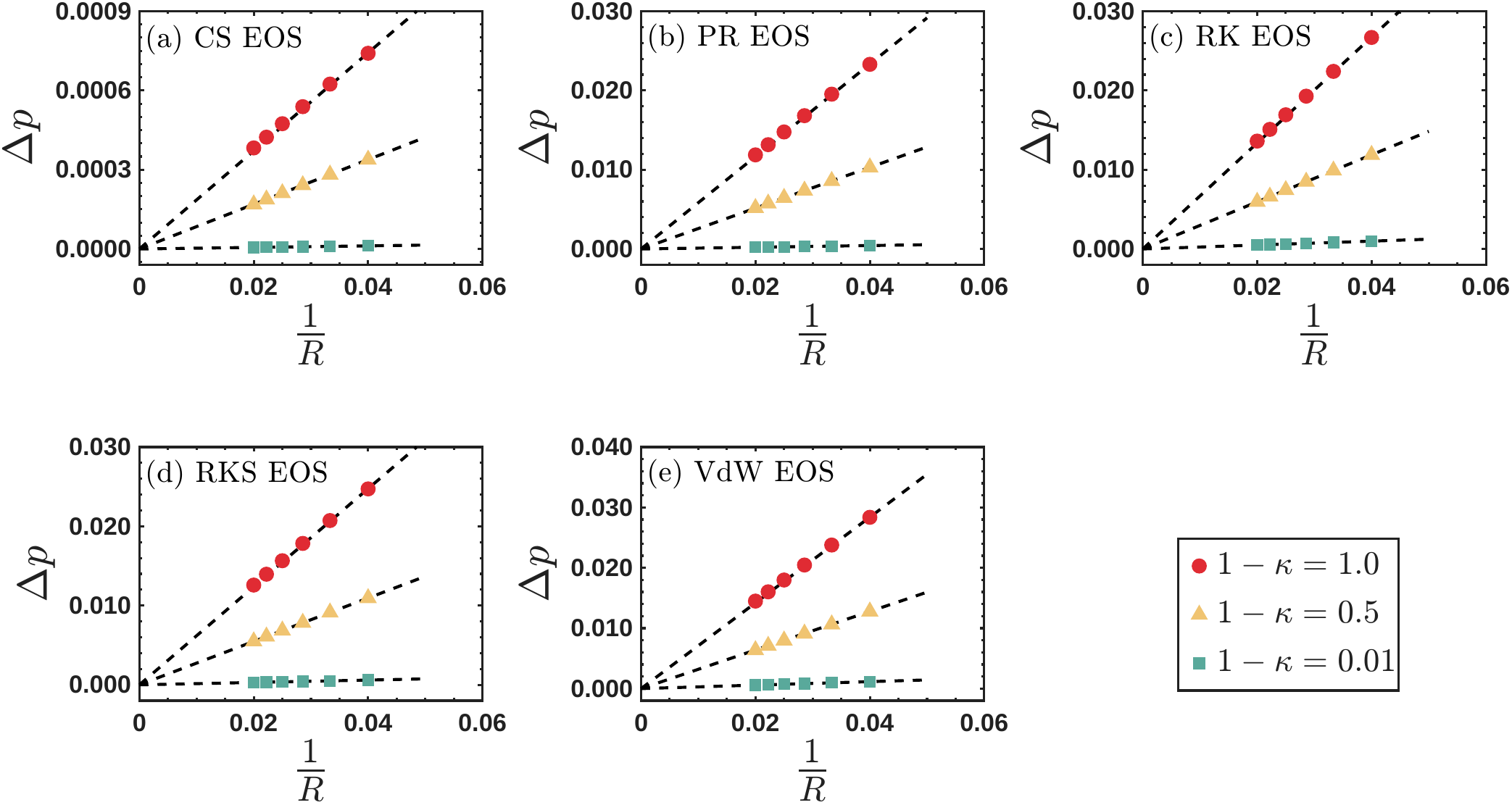}
    \caption{The relationship between pressure differential and droplet curvature (1/$R$) plotted for various values of $\kappa$, shown separately for (a) Carnahan-Starling (CS), (b) Peng-Robinson (PR), (c) Redlich-Kwong (RK), (d) Redlich-Kwong-Soave (RKS) and (e) van der Waals (VdW) equation of state at $\tau_v = 1.0$ and $T_r = 0.8$. All values are in lattice units.}
    \label{fig:eos_laplace}
\end{figure}

To validate the numerical implementation described in \ref{multiphase-implementation} against Laplace's law, 2D and 3D simulations of a circular droplet are conducted using the D2Q9 and D3Q19 lattices, respectively. Laplace's law defines the relationship between the pressure difference inside and outside the droplet ($\Delta p$) and inverse droplet radius ($1/R$):
\begin{equation}
    \Delta p = \frac{\sigma}{R}
\end{equation}
where $\sigma$ is the surface tension constant. The simulations are performed in a domain size of $200^2$ for 2D and $200^3$ for 3D, with periodic boundary conditions applied throughout. A circular droplet is placed at the center, with radius varying between 25 and 50 lattice units with increments of 5 lattice units (Fig. \ref{fig:droplet_test}). The reduced temperature $T_r$ is set as 0.8 for all EOS considered, and the density field is initialized using Eq. \ref{eq:laplace_init_density} with width $w = 3$ lattice units. To adjust surface tension, the MRT collision model is used. Similar to the previous case, the parameters for each of equation of state are taken from \citeA{yuan_equations_2006} and the values of $\alpha^k$ and $A_{kk'}$ in Eq. \ref{eq:software_ff_force_mcmp} are adjusted for each EOS to ensure that the mechanical stability condition is satisfied. Figure \ref{fig:eos_laplace} illustrates that the computed pressure difference across the interface is proportional to the inverse radius of the droplet for all equations of state considered, thus verifying the Laplace law. 

Moreover, the coexistence densities obtained for different surface tension adjustment coefficients $\kappa$ for the identical testing case are tabulated in Table \ref{table:coex_dens}. The results reveal that the computed densities for all EOS do not vary with $\kappa$ and are in close agreement with the predictions from the Maxwell construction. Furthermore, the surface tension values listed in Table \ref{table:eos_surface_tension} indicate that surface tension can be modified by factors of 110, 58, 100, 92, and 115 for the CS, PR, RK, RKS, and VdW equations of state, respectively, without altering the density ratio proving that the pressure tensor modification methods allows independent adjustment of surface tension without affecting the liquid-vapor coexistence densities unlike the original Shan-Chen method.
\begin{table}[!ht]
\caption{Comparison between simulated and Maxwell coexistence densities for different surface tension adjustment coefficient $\kappa$ at $T_r = 0.8$. All values are in lattice units.}
\begin{center}
\begin{tabular}{ p{1.5cm} p{2.5cm} p{2.5cm} p{2.5cm} p{2cm} }
\hline
\multicolumn{5}{c}{$\rho_g/\rho_l$}\\
\hline
EOS & $1 - \kappa = 1.0$ & $1 - \kappa = 0.5$ & $1 - \kappa = 0.01$ & Maxwell Construction\\
\hline
CS & 0.0217/0.3091 & 0.0217/0.3088 & 0.0217/0.3072 & 0.0217/0.3071\\
PR & 0.1971/7.2063 & 0.1971/7.2063 & 0.1971/7.2063 & 0.1970/7.2040 \\
RK & 0.3425/6.6734 & 0.3426/6.6734 & 0.3427/6.6734 & 0.3428/6.6252 \\
RKS & 0.2210/7.0692 & 0.2210/7.0695 & 0.2210/7.0697 & 0.2208/7.0693 \\ 
VdW & 0.8389/6.7635 & 0.8381/6.7605 & 0.8386/6.7681 & 0.8388/6.7644 \\
\hline
\end{tabular}
\end{center}
\label{table:coex_dens}
\end{table}
\begin{table}[!ht]
\caption{Surface tension values obtained using the present method for different surface tension adjustment coefficient $\kappa$ at $T_r = 0.8$. All values are in lattice units.}
\begin{center}
\begin{tabular}{ p{1.5cm} p{2.5cm} p{2.5cm} p{2.5cm} }
\hline
\multicolumn{4}{c}{Surface tension}\\
\hline
EOS & $1 - \kappa = 1.0$ & $1 - \kappa = 0.5$ & $1 - \kappa = 0.01$ \\
\hline
CS & 0.0186 & 0.0085 & 0.00017 \\
PR & 0.5831 & 0.2574 & 0.0101 \\
RK & 0.6684 & 0.2969 & 0.00681 \\
RKS & 0.6189 & 0.2738 & 0.00675 \\ 
VdW & 0.7099 & 0.3194 & 0.00617 \\
\hline
\end{tabular}
\end{center}
\label{table:eos_surface_tension}
\end{table}
\subsection{Capillary Flows}
\subsubsection{Capillary rise in a tube}\label{washburn}
\begin{figure}[!ht]
    \centering
    \subfloat[][]{
    \includegraphics[scale=0.2]{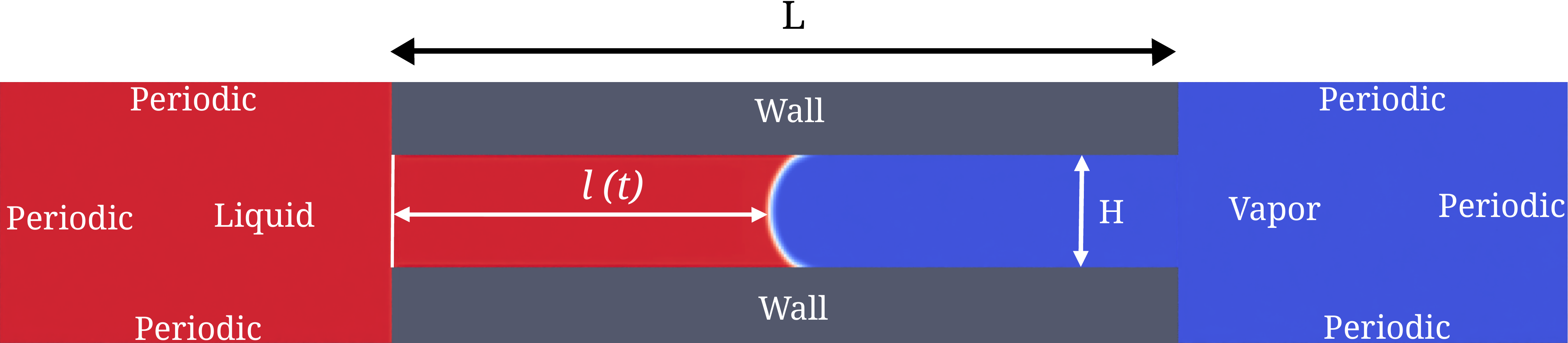}
    \label{fig:lucas_washburn_schematic}} \\
    \subfloat[][]{
    \includegraphics[width=7cm,height=6cm]{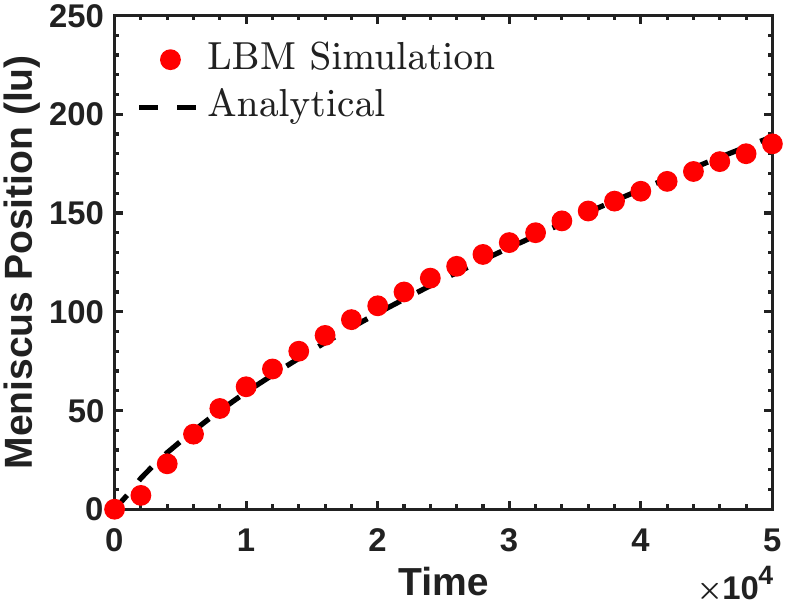}
    \label{fig:lucas_washburn}}
\caption{(a) Schematic diagram of capillary rise process. (b) Comparison between imbibition lengths obtained from simulation and analytical expression (in lattice units).}
\end{figure}

During spontaneous imbibition in channels, the wetting fluid advances by displacing the non-wetting phase, and this process can be classified into liquid-liquid or gas-liquid drainage. Gas-liquid drainage in particular has been the focus of extensive study because of its importance in applications such as fracturing-fluid invasion in unconventional reservoirs, $\mathrm{CO_2}$ storage, and aqueous-phase transport in underground gas storage reservoirs \cite{yang_recent_2023}. A 2D single-component simulation is performed using the MRT collision model and D2Q9 lattice on a 600 × 100 domain. A channel with length ($L$) of 300 lattice units and width of 28 lattice units is set at the position of $150 < x < 450$, with static contact angle $\theta = 19^\circ$. No-slip (bounceback) boundary condition is applied at the channels, while a periodic boundary condition is applied at all other peripheral nodes. The simulation is initialized with positions $x < 150$ occupied by liquid (wetting phase), and the rest occupied by the gas phase. VanderWaal equation of state with parameters $a = 9/49$, $b = 2/21$, and $R = 1$ is used to model the single-component multiphase system at reduced temperature $T_r = 0.7$. The surface tension adjustment factor is set as $\kappa = 0.5$. Considering capillary, inertia, and viscous forces, the meniscus position $l$ at time $t$ for a fluid undergoing horizontal imbibition can be determined using the following equation \cite{yang_characterizing_2025}:
\begin{equation}
    \frac{2\sigma cos(\theta)}{H} - [\rho_l l + (L - l)\rho_v]\frac{d^2 l}{d t^2} - [\rho_l l + (L - l)\rho_v](\frac{d l}{d t})^2 - \frac{12}{H^2}[\mu_l l + (L - l)\mu_v] = 0
    \label{eq:lucas_washburn_ode}
\end{equation}
where $\sigma$, $H$, $\rho$, $\mu$ are the surface tension coefficient, width of the capillary, density, and dynamic viscosity, respectively (Fig. \ref{fig:lucas_washburn_schematic}; subscripts $l$ and $v$ denote respective values for liquid and vapor phase.) Eq. \ref{eq:lucas_washburn_ode} can be simplified by ignoring gas-phase viscosity, inertial forces, and dynamic wetting effects to get the original Lucas-Washburn equation \cite{washburn_dynamics_1921}. Although the Lucas-Washburn equation closes matches macroscopic experimental results, at micro-scale and nano-scale, various fluid properties such as viscosity and contact angle are not constant and hence, the Lucas-Washburn equation leads to an overprediction of meniscus position as confirmed in several experimental studies \cite{siebold_effect_2000, heshmati_experimental_2014}. Among these parameters, the most noticeable impact on meniscus height prediction comes from using a static contact angle, instead of a dynamic contact angle \cite{hamraoui_can_2000}. Thus, in this paper, Eq. \ref{eq:lucas_washburn_ode} is solved using fourth-order explicit Runge-Kutta scheme with initial conditions $l(0) = 0$ and $\frac{dl}{dt}|_{t=0} = 0$\cite{wolf_capillary_2010} and the dynamic contact angle $\theta_d$ is measured during the simulation using the height of the bent liquid surface $h_m$ \cite{yang_characterizing_2025}:
\begin{equation}
    \theta_d = \mathrm{cos}^{-1} (\frac{4 h_m H}{H^2 + 4h_m^2})
\end{equation}
Fig. \ref{fig:lucas_washburn} plots the meniscus positions obtained from the LB simulation against those obtained from the analytical expression, showing close agreement between the two. 
\subsubsection{Drainage regimes in a channel}\label{channel_drainage}
\begin{figure}[!ht]
    \centering    \noindent\includegraphics[scale=0.5]{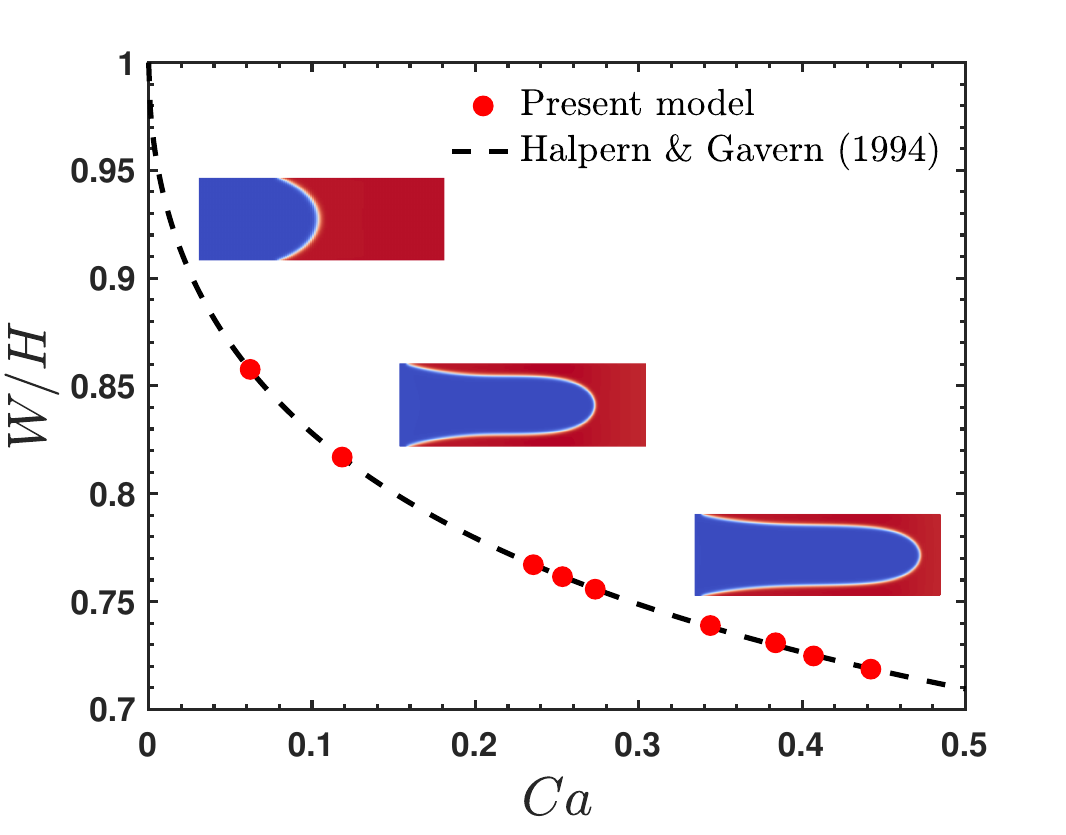}
    \caption{Comparison of the relative finger width (finger width normalized by channel width) obtained using the present method with the empirical relation from \protect\citeA{halpern_boundary_1994} for density ratio of 1.0. The inset figure illustrates finger shapes at different $Ca$ values, showing stable displacement at low $Ca$}
    \label{fig:capillary_fingering}
\end{figure}

This section benchmarks the morphology of an invading non-wetting front (i.e., drainage) within a capillary channel for the case of immiscible displacement agsainst the empirical model developed in \citeA{halpern_boundary_1994}. Depending on the dominant mechanism, the displacement can be classified as either capillary fingering, viscous fingering, or stable displacement, and the influence of capillary and viscous forces is quantified using two dimensionless numbers: the capillary number ($Ca$) and the viscosity ratio ($M$) \cite{lenormand_numerical_1988}. 
A 2D multicomponent fluid simulation is performed for two fluids in a channel using the D2Q9 lattice and MRT collision model. Component 1 occupies the left half of the channel, while component 2 occupies the right half. Initially, a small fraction of each fluid component is introduced into the bulk of the other to establish miscibility, with the exact value determined by the strength of inter-component interactions. It is crucial to ensure that each component maintains a non-zero density throughout the domain to prevent division by zero during velocity computation (Eq. \ref{eq:macroeq}). In this benchmark, the fraction is fixed at 3\%, following the approach of \citeA{kang_immiscible_2004}. The intra-component interaction strength is set to a small negative value of -0.027, while the inter-component interaction strength is assigned a large positive value of 0.57. These values are tuned to minimize the thickness of the interface between component 1 and component 2, thereby promoting immiscibility. The channel inlet and outlets are set as periodic boundaries, while the channel walls are set as no-slip boundaries. A body force is applied along the channel to drive the fluids. The simulations are performed until the breakthrough of component 1 is achieved. The interaction forces are computed using the original Shan-Chen scheme by setting $A_{kk'} = 0$ in Eq. \ref{eq:software_ff_force_mcmp}, and the pseudopotential computation function in the code is modified to use the component density as pseudopotential ($\psi^k = \rho^k$). The simulations are performed for the case of both fluids being neutrally-wet (contact angle = $90^\circ$, $\phi = 1.0$, and $\Delta \rho = 0.0$ in equation \ref{eq:improved_virtual_density}). The surface tension coefficient used for computing $Ca$ is determined using Laplace's law, as outlined in \ref{laplace_law}. 
The channel simulations are performed using a $500 \times 75$ lattice, and the prescribed force along the flow direction is varied to achieve various capillary numbers. The interface is defined at the location where both components have identical density, and the maximum finger width is determined using the ImageJ Fiji software \cite{schindelin_fiji_2012}. The relative width of the finger (finger width/channel width) obtained in each case is plotted as a function of $Ca$ in Fig. \ref{fig:capillary_fingering}, and these results agree with the empirical model developed in \citeA{halpern_boundary_1994}'s work.
\subsubsection{Relative permeability in layered cocurrent flow}\label{cocurrent}
\begin{figure}[!ht]
\centering
\subfloat[][]{
\includegraphics[scale=0.49]{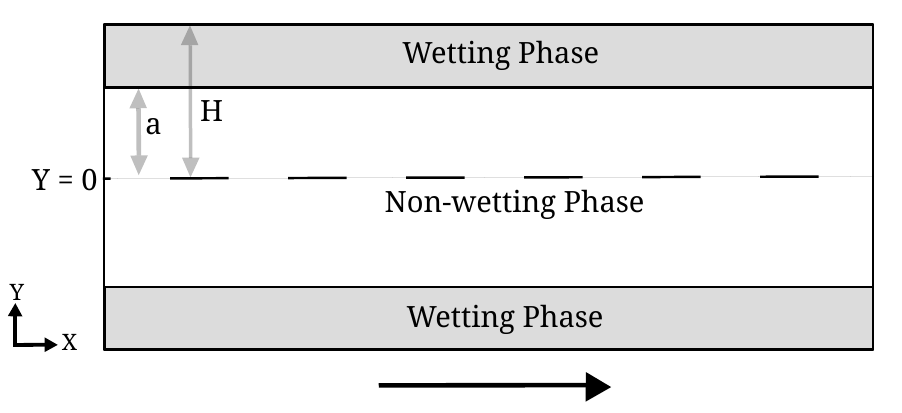}
\label{fig:cocurrent_domain}}
\subfloat[][]{
\includegraphics[scale=0.34]{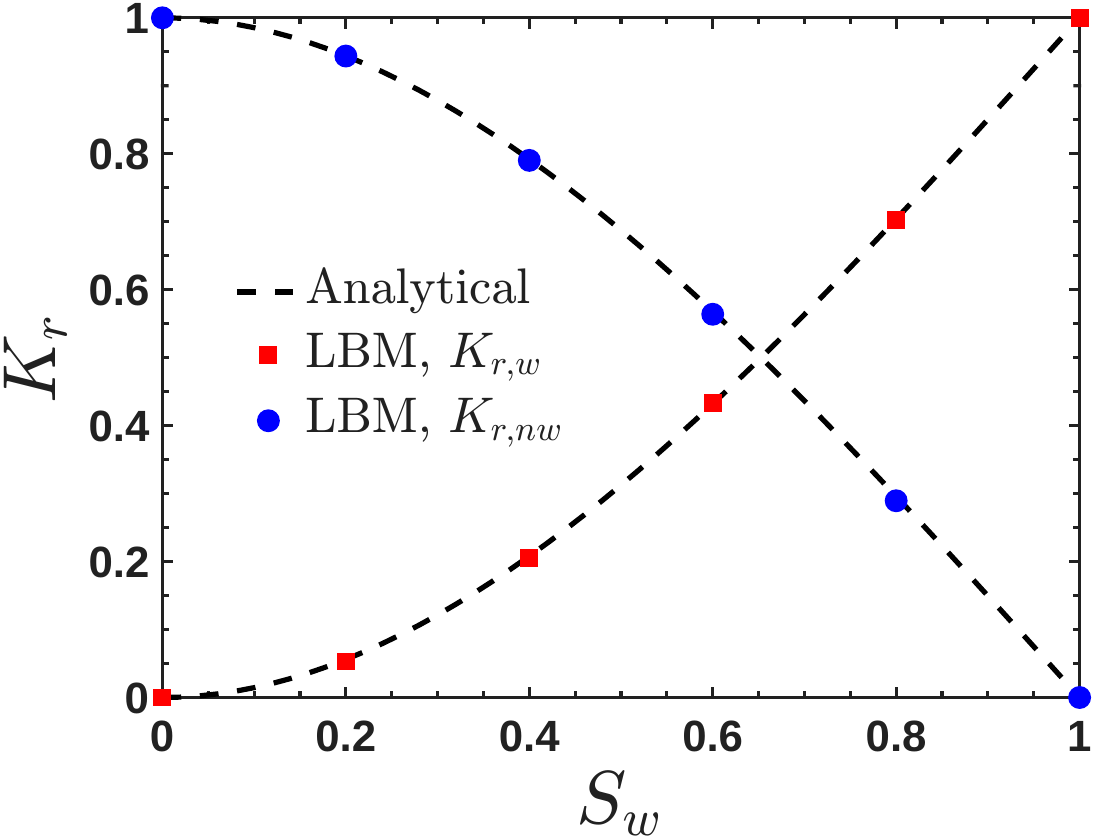}}
\caption{(a) Schematic of two-phase cocurrent flow in a channel. (b) Comparison of simulated and analytical relative permeability curves for the wetting phase $(K_{r,w})$ and non-wetting phase $(K_{r,nw})$ in cocurrent flow with a viscosity ratio of $M = 1.0$.}
\label{fig:cocurrent_rel_perm}
\end{figure}

The cocurrent flow of two components in a channel is illustrated in Fig. \ref{fig:cocurrent_domain}. The central region contains the non-wetting phase, while the surrounding area is occupied by the wetting phase. The flow is driven by a body force, and the relative permeability ($K_r$) of each phase is determined as the ratio of its flow rate in cocurrent flow to its flow rate in a single-phase simulation:
\begin{equation}
    \begin{gathered}
        K_{r,w} = \frac{\int_{|y| = 0}^{a} u_{nw} dy}{\int_{|y| = 0}^H u_{nw} dy} \\
        K_{r,w} = \frac{\int_{|y| = a}^{H} u_{w} dy}{\int_{|y| = 0}^H u_{w} dy}
    \end{gathered}
\end{equation}

The steady state expression for the relative permeability ($K_r$) can be obtained analytically, assuming laminar flow and a sharp interface between the components \cite{porter_multicomponent_2012}:
\begin{equation}\label{eq:channel_rel_perm}
    \begin{gathered}
        K_{r, w} = \frac{1}{2}S_{w}^2(3 - S_w) \\
        K_{r, nw} = (1 - S_w)[\frac{3}{2}M + (1 - S_w)^2(1 - \frac{3}{2}M)]
    \end{gathered}
\end{equation}
where $S_w$ is the wetting phase saturation, defined as the ratio of wetting phase width to channel width ($a/H$), and $M$ is the dynamic viscosity ratio.

A 2D multicomponent fluid simulation is performed for flow in a channel using D2Q9 lattice and the MRT collision model. The channel walls are set as no-slip boundaries, while periodic boundary conditions were applied at the inlet and outlet. Using density as pseudopotential ($\psi^k = \rho^k$), the inter- and intra-component interaction forces are tuned to ensure that the interface thickness between two components is less than or equal to 4 lattice units. Fig. \ref{fig:cocurrent_rel_perm} shows a close agreement between the simulated relative permeability values with the equation \ref{eq:channel_rel_perm} analytical results for viscosity ratio $M = 1$.
\subsection{Porous Media Flows}\label{applications}
\subsubsection{Permeability estimation for a synthetic porous media}\label{permeability_estimation}
JAX-LaB is utilized to determine the permeability of a given porous geometry. The geometry used here is obtained from the Digital Rocks Portal \cite{santos_3d_2022, drp_geometry} (geometry: \verb|374_05_03_256|) and represents a $256^3$ randomly placed overlapping sphere pack with an embedded fracture, having porosity 0.1962. To compute permeability, a pressure gradient of $0.01$ in lattice units is applied along the flow direction ($x$-direction), while a no-slip boundary condition is imposed along the solid nodes of the geometry, $y$ and $z$ direction faces. The domain is initialized with zero velocity everywhere and a linear pressure profile along the $x$ direction. The simulation is run until a steady state is achieved (50,000 lattice timesteps). Plugging the values of the steady-state mean velocity obtained from simulation and the applied pressure gradient into Darcy's law, the permeability of the model is determined to be 2.003 Darcy, which falls within the expected range of 2-3 Darcy \cite{drp_geometry}.
\begin{figure}[!ht]
    \centering
    \noindent\includegraphics[scale=0.14]{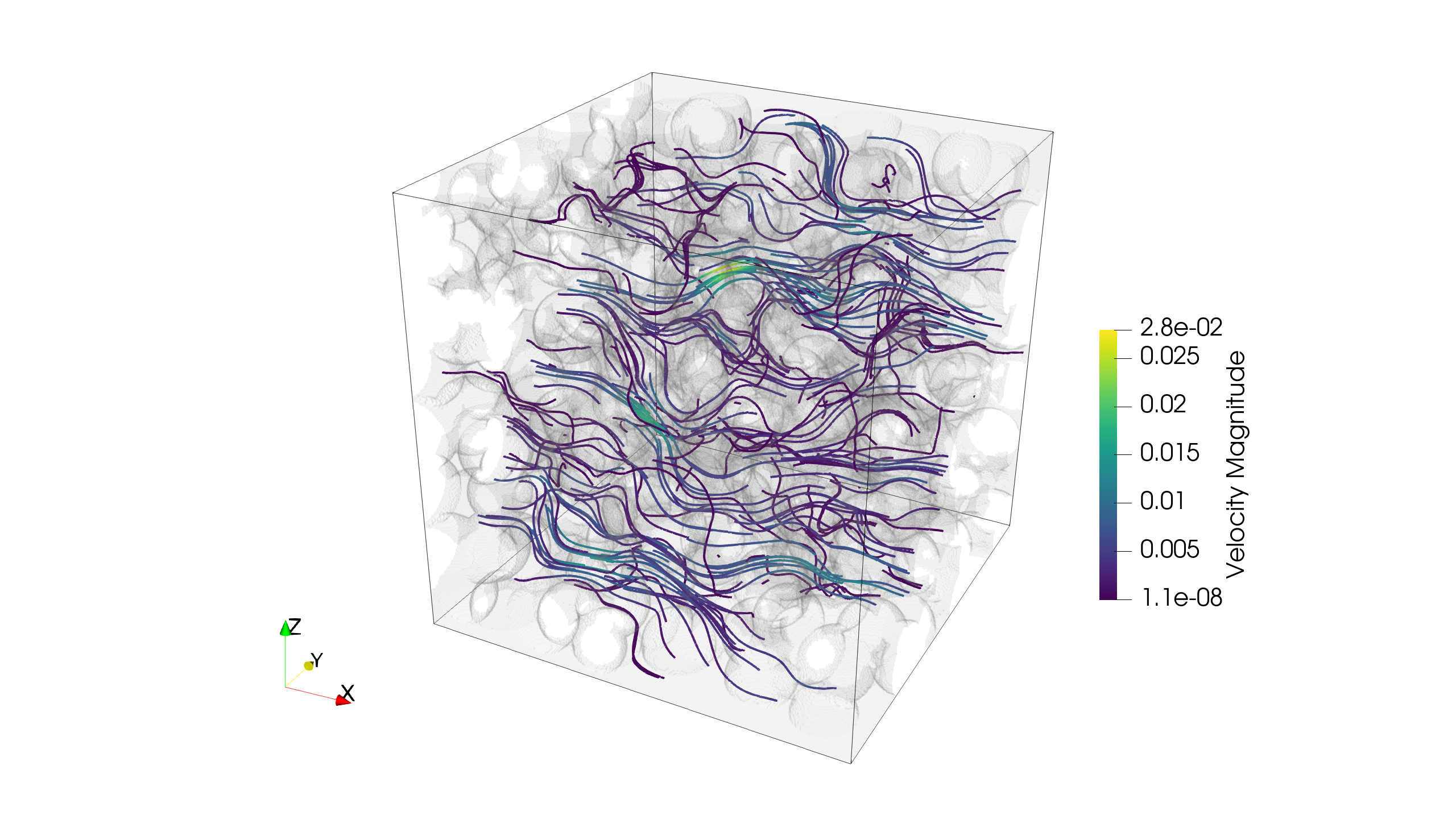}
    \caption{Streamlines for the steady-state solution in a core comprising of randomly placed overlapping spheres and an embedded synthetic fracture obtained from the Digital Rocks Portal.}
    \label{fig:single_phase_permeability}
\end{figure}
\subsubsection{Drainage simulation in Fontainebleau sandstone}
\begin{figure}[H]
\centering
\subfloat[][]{
    \centering
    \includegraphics[scale=0.15]{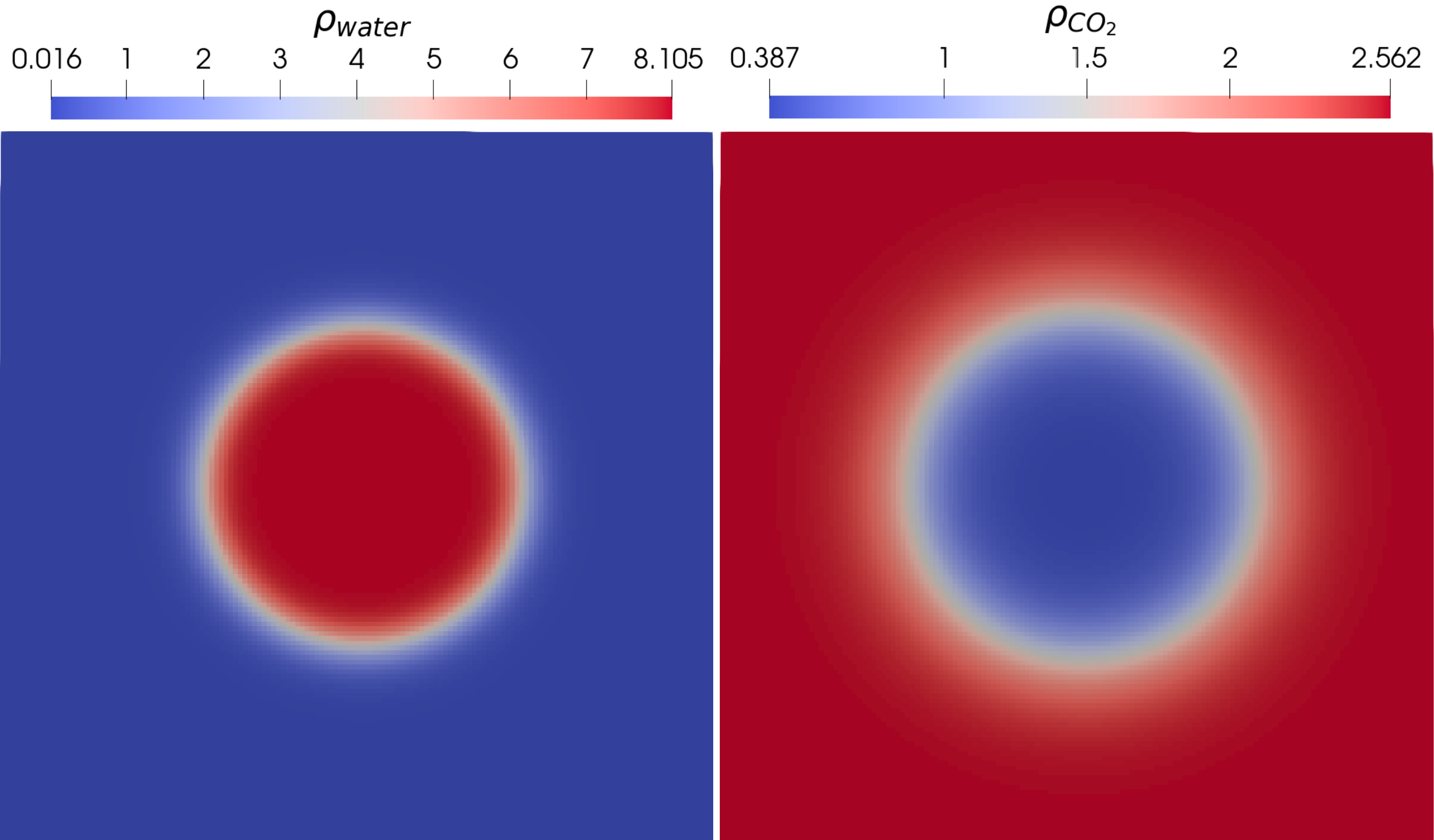}
    \label{fig:multicomponent_droplet}
}
\subfloat[][]{
    \centering
    \includegraphics[scale=0.14]{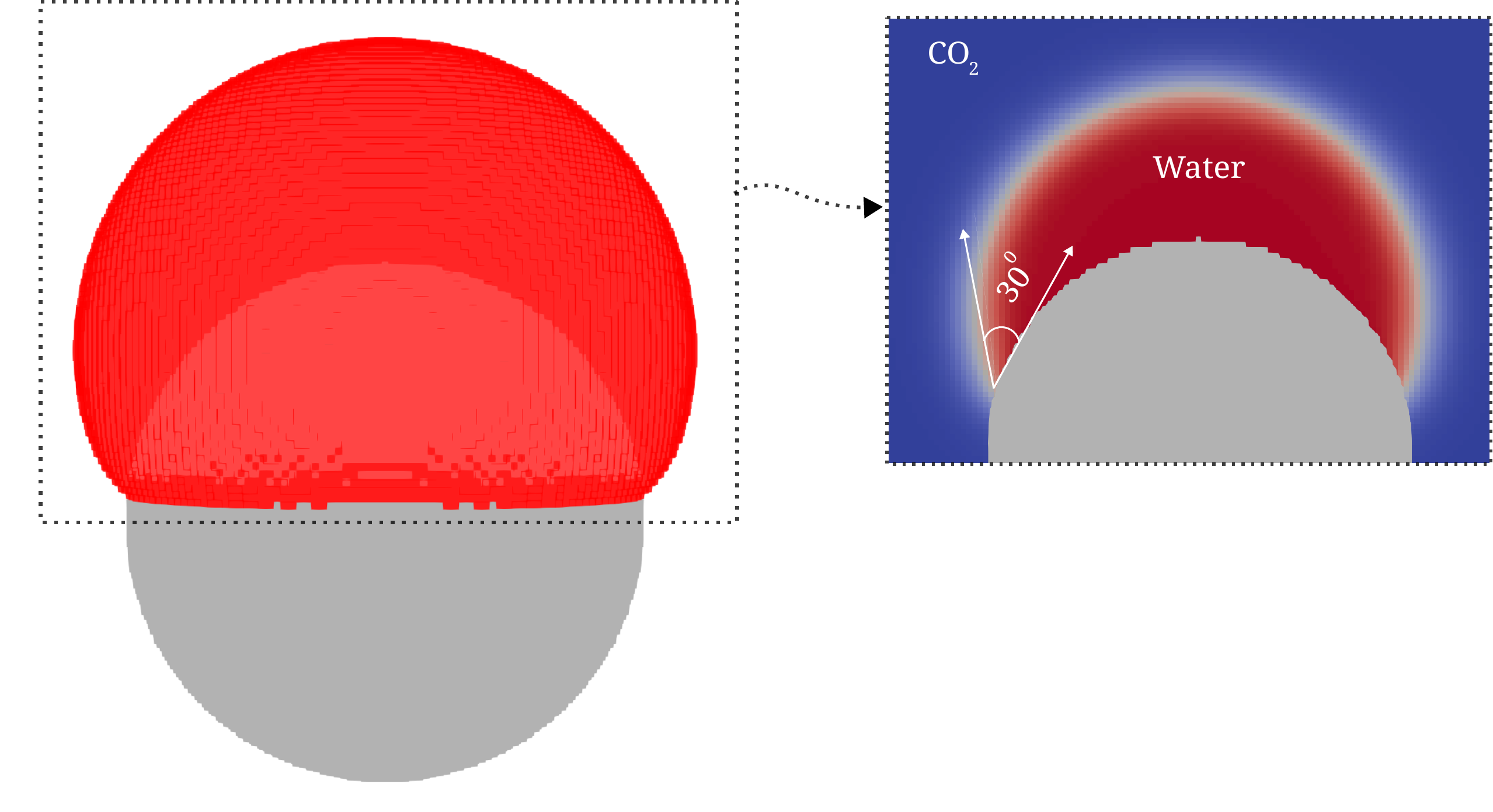}
    \label{fig:multicomponent_droplet_curved}
}
\caption{Three-dimensional multi-component droplet simulation (with density values shown in lattice units): (a) for determining fluid-fluid interaction strengths and (b) for evaluating fluid wettability parameters on a curved surface. A density threshold is used to isolate the droplet (red) from the background for visualization, and the contact angle formed on spherical geometry (grey) is measured at steady state. The density profiles are shown for a slice taken at droplet center (\protect$y=75$).}
\label{fig:multi_component_drainage_parameter_tuning}
\end{figure}

This section models a two-component flow of water and supercritical $\mathrm{CO_2}$ through a Fontainebleau sandstone obtained from the Digital Rocks portal \cite{drp_geometry} and showcases the use of JAX-LaB for modeling multicomponent, multiphase flow in porous media, with applications in $\mathrm{CO_2}$ sequestration \cite{ashirbekov_equation_2021}, enhanced oil recovery \cite{xu_pore-scale_2020}, evaporation \cite{fei_pore-scale_2022} and beyond. It also describes the procedure used for setting the inter-component strengths and wettability parameters in multi-component simulations. The parameters used in equations \ref{eq:software_ff_force_mcmp}, \ref{eq:li_surface_adjust}, and \ref{eq:improved_virtual_density}, which govern inter-fluid interaction and wettability, are calibrated using a two-component droplet simulation ($150^3$ domain, D3Q19 lattice and MRT collision model). The simulation is initialized with a central water droplet surrounded by supercritical $\mathrm{CO_2}$. The initial density distribution and mutual solubility of the fluids are tabulated in table \ref{table:eos_brine_co2}, and are set following \citeA{tang_multi-component_2024}, corresponding to fluid properties at 20 MPa and $200^\circ$C. We employ the Peng-Robinson equation of state in this supercritical regime because its predictions align well with experimental observations \cite{ashirbekov_equation_2021, tang_multi-component_2024}. The inter-fluid interaction coefficient ($g_{kk'}$) is chosen to be small and positive (weakly repulsive), while the intra-fluid coefficient ($g_{kk}$) is set to -1 (indicating strong cohesion), and the modification coefficient $A_{kk}$ is set to 0. Relaxation times are selected to maintain a viscosity ratio of $\frac{\tau_{H_2O}}{\tau_{CO_2}} = 1.43$, representing the viscosity contrast between water and sc$\mathrm{CO_2}$, assuming identical viscosities for liquid and vapor states of water. Surface tension of this system is calculated using Laplace's law (modification coefficient $\kappa_{\mathrm{CO_2}} = 0.2$). The steady-state densities obtained from the droplet simulation lead to a density ratio of 506. The wettability of the system is characterized by the steady-state contact angle formed by a water droplet suspended in sc$\mathrm{CO_2}$ on a spherical surface. Parameters $\phi$ and $\Delta \rho$ are then tuned to produce a contact angle of $30^\circ$, following \citeA{chen_inertial_2019}. The consolidated results from these simulations, used for parameter estimation, are presented in Fig. \ref{fig:multicomponent_droplet} and \ref{fig:multicomponent_droplet_curved}. These parameters are subsequently applied in the pore-scale simulations. 

\begin{figure}[H]
\centering
\includegraphics[scale=0.14]{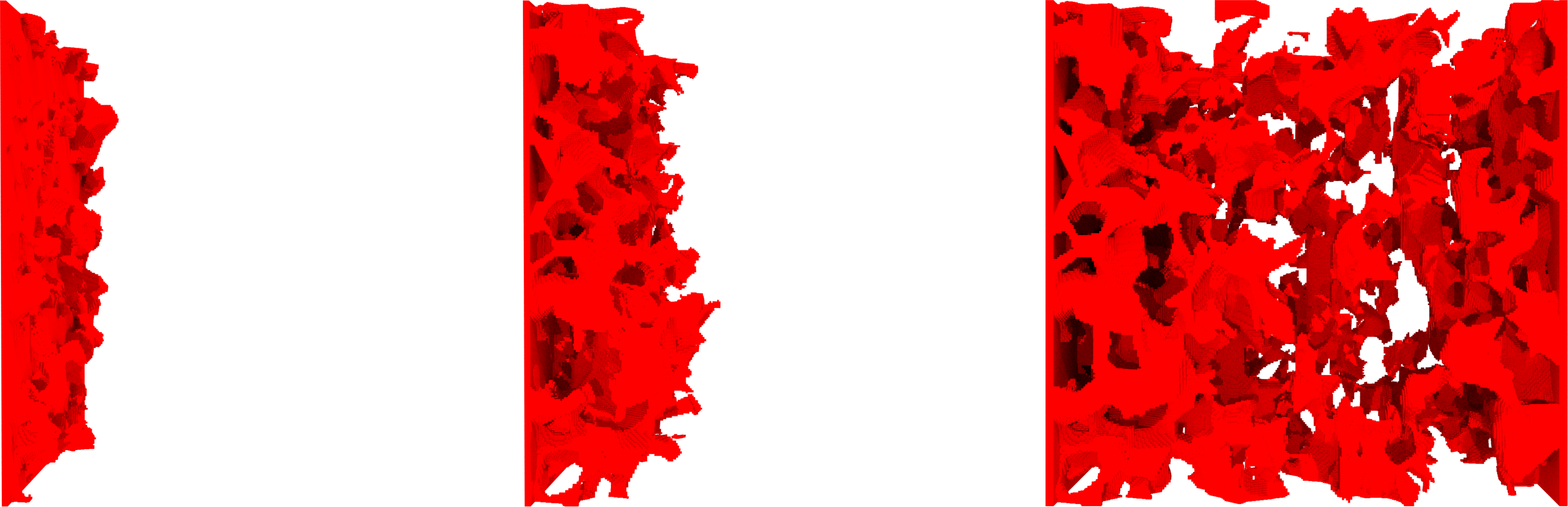}
\label{fig:drainage}
\caption{Snapshots of invading component (sc\protect$\mathrm{CO_2}$) distributions for simulation of drainage in a water-saturated Fontainebleau sandstone.}
\label{fig:multi_component_drainage}
\end{figure}

The top, bottom, front, and back faces of the $256^3$ Fontainebleau sandstone (with porosity of 0.1938) domain are set as periodic boundaries, and the solid nodes in the domain are set as no-slip boundaries using the bounceback scheme. Buffer layers are added to the domain at the inlet and outlet to prevent end effects at the domain boundaries. Initially, sc$\mathrm{CO_2}$ is assumed to fully occupy the inlet buffer layer, and it is driven into the system using body force. For the outlet, a convective boundary condition \cite{lou_evaluation_2013} based on average velocity is used. The thickness of the input buffer layer is set such that the amount of sc$\mathrm{CO_2}$ present in it is enough to fill all the pores in the domain. The snapshots from the drainage simulation at various stages of saturation are shown in Fig. \ref{fig:multi_component_drainage}.
\begin{table}[!ht]
\caption{Parameters of Peng-Robinson equation of state for carbon dioxide and water in lattice units from \protect\citeA{tang_multi-component_2024}.}
\begin{center}
\begin{tabular}{ p{2.4cm} p{1cm} p{1cm} p{1cm} p{0.2cm}}
\hline
Component & a & b & $\omega$ & R \\
\hline
Water & 0.0204 & 0.09523 & 0.344 & 1\\
Carbon dioxide & 0.01348 & 0.13385 & 0.22491 & 1\\
\hline
\end{tabular}
\end{center}
\label{table:eos_brine_co2}
\end{table}
\subsubsection{Characteristic curves for a sphere pack}
\begin{figure}[H]
\centering
\subfloat[][]{
    \centering
    \includegraphics[scale=0.57]{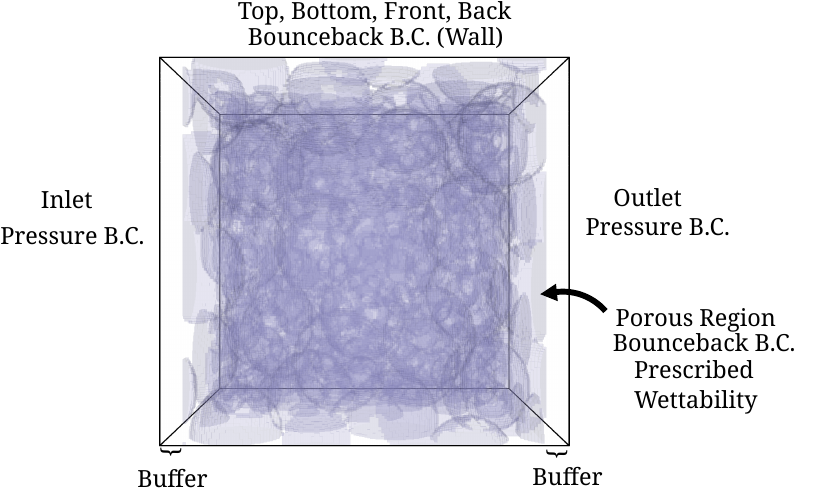}
    \label{fig:swcc_domain}
}
\subfloat[][]{
    \centering
    \includegraphics[width=6.3cm, height=5.3cm]{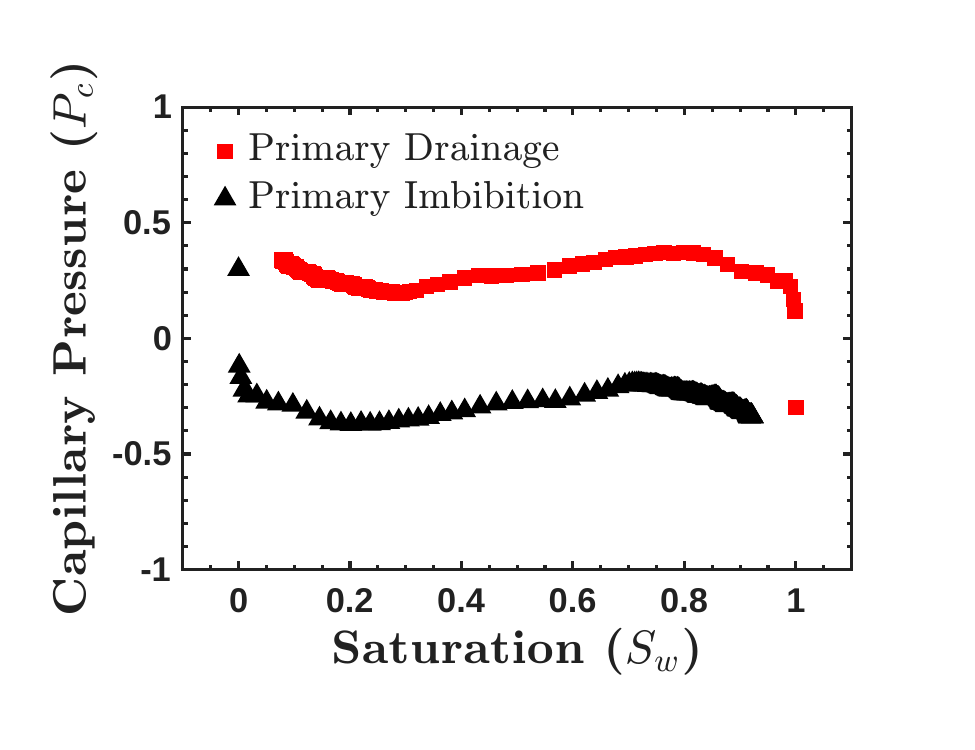}
    \label{fig:swcc}
}
\vspace{1em}
\subfloat[]{
\includegraphics[scale=0.07]{swcc_drainage.pdf}
\label{fig:swcc_drainage}}
\caption{(a) Schematic of simulation domain used to obtain drainage/imbibition curves for a sphere pack. (b) Characteristic curves obtained from the drainage/imbibition process. (c) Distributions of non-wetting component during a primary drainage simulation for the synthetic sphere pack at various timesteps. The translucent grey region represents the solid phase, and the wetting component is not shown for visual clarity. The images are generated in situ from JAX arrays on GPU using the PhantomGaze library \protect\cite{phantomgaze}.}
\label{fig:multi_component_swcc}
\end{figure}

\begin{figure}[H]
\centering
\subfloat[\label{fig:ml-combined-results}]{
    \noindent\includegraphics[width=13.4cm]{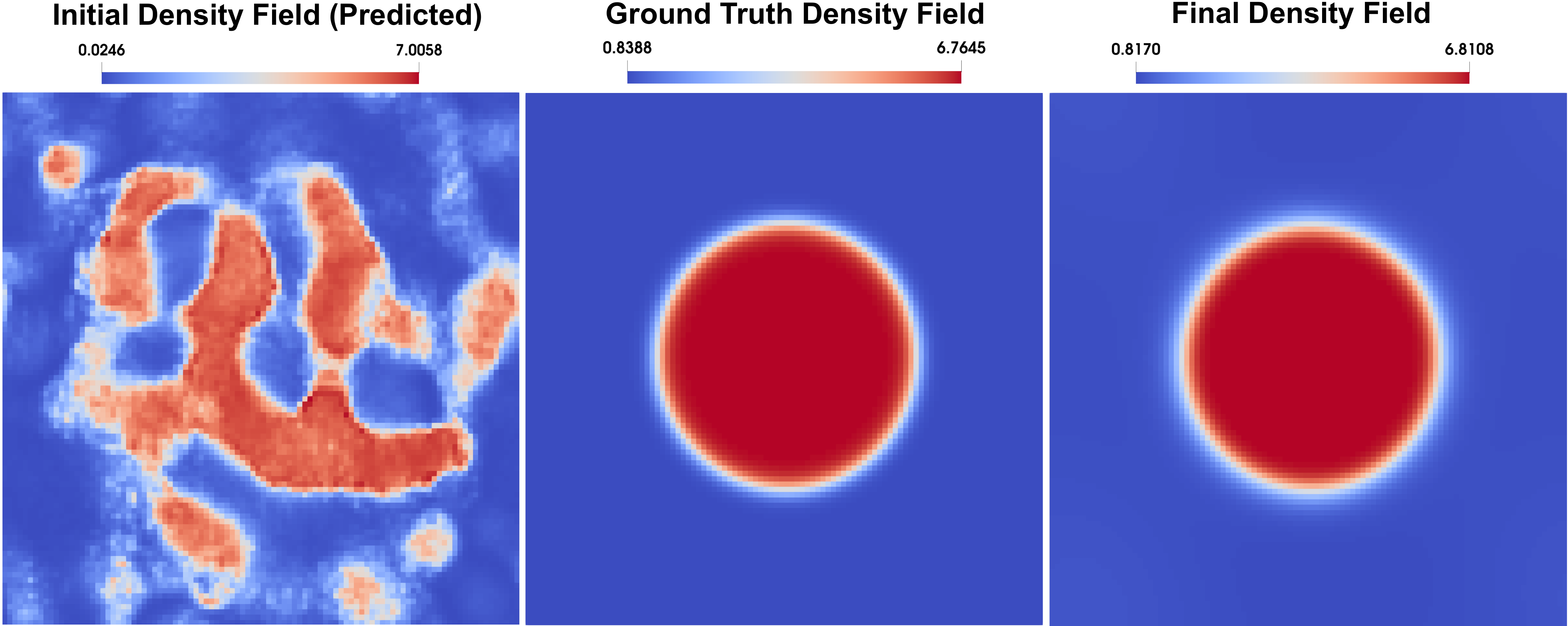}
}
\vspace{1em}
\centering
\subfloat[\label{fig:ml-predicted-horizon}]{
    \noindent\includegraphics[scale=0.134]{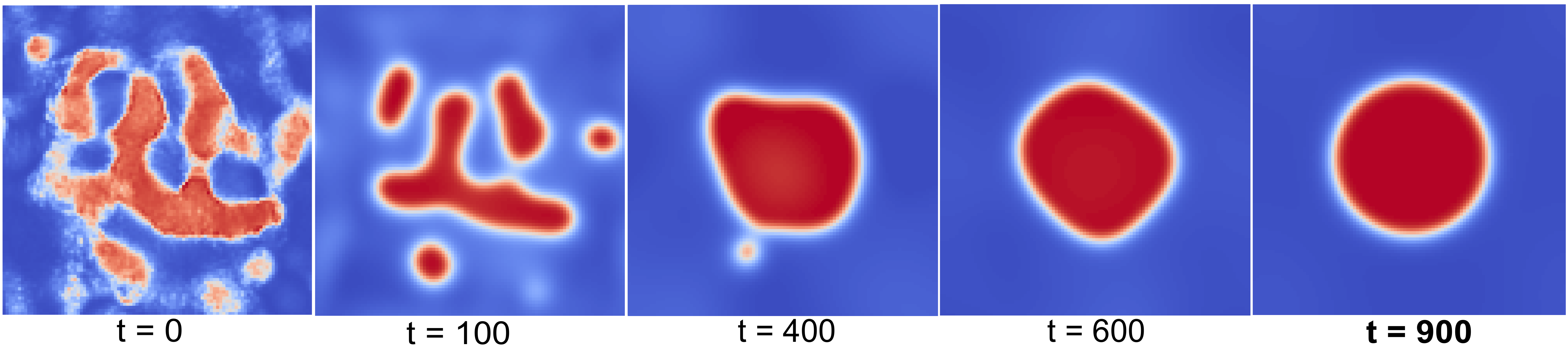}
}
\caption{\protect\add[PP]{(a) Illustration of the initial density field predicted by the neural network, the target (ground truth) density field, and the resulting density field after 900 iterations. (b) Temporal evolution of the density field showing the initial distribution, and the formation of the droplet at $t = 900$.}}
\label{fig:consolidated}
\end{figure}

This section illustrates the application of JAX-LaB for generating capillary pressure ($P_c$) v/s wetting component saturation ($S_w$) characteristic curves for a $256^3$ synthetic sphere pack (geometry: \verb|374_05_03_256|) with a porosity of 0.381 sourced from the Digital Rocks Portal \cite{drp_geometry}. It highlights JAX-LaB's capabilities for simulating multicomponent, multiphase flow in porous media, as well as its in-situ measurement and visualization features that eliminate the need for post-processing. A 3D multiphase simulation is conducted using the BGK collision model, D3Q19 lattice, and two fluid components. Following the methodology of \citeA{galindo-torres_boundary_2016}, both components are assigned the same density. Wettability effects are introduced by tuning the parameters in Eq. \ref{eq:improved_virtual_density} to yield a contact angle of $45^\circ$ for the wetting component. The simulation domain is depicted in Fig. \ref{fig:swcc_domain}. Buffer layers, each 8 lattice units thick, are added near the inlet and outlet, with wall boundary conditions applied to all other faces and the porous matrix. Constant pressure boundary condition is applied at the inlet and outlet to drive the fluids. Although flux boundary conditions have been employed in previous studies for capillary pressure-saturation measurements \cite{porter_lattice-boltzmann_2009, galindo-torres_boundary_2016}, we adopt pressure boundary conditions here because they more closely mirror the experimental procedures used to derive characteristic curves \cite{porter_lattice-boltzmann_2009} and enable efficient capture of complete drainage and imbibition behavior in a single simulation, starting from either a fully saturated or fully dry state. To obtain data points along the curve, $P_c$ and $S_w$ are computed after every 1000 lattice timesteps, with each simulation running for a total of 250,000 timesteps. Since both components remain present throughout the domain, a threshold-based segmentation is applied: lattice sites where the non-wetting component density equals or exceeds that of the wetting component are classified as non-wetting \cite{porter_lattice-boltzmann_2009, galindo-torres_boundary_2016}. $P_c$ is computed using the average pressure of each component within its respective regions after segmentation, while $S_w$ is calculated as the fraction of sites containing the wetting component. Fluid distributions within the porous domain are visualized in situ during the simulation using JAX arrays stored on GPU through the PhantomGaze library \cite{phantomgaze}, enabling real-time monitoring without post-processing. Figure \ref{fig:swcc_drainage} demonstrates this capability by showing the evolution of the non-wetting component at selected timesteps during the drainage simulation.
\add[PP]{\protect\section{Physics-based Machine Learning}
The native differentiability of XLB and its applications in scientific machine learning have been thoroughly discussed by \cite{ataei_xlb_2024}, who demonstrated that a unified computational graph for both forward and backward passes enables seamless integration of neural components within physical simulations. This framework allows machine learning models to be incorporated seamlessly at multiple stages of the simulation workflow. In this section, we demonstrate how JAX-LaB extends these capabilities to multiphase flow systems by addressing an inverse flow control problem, where automatic differentiation is employed during the neural network’s backpropagation (Fig. \ref{fig:ml-workflow}). For this demonstration, we use the JAX-based Equinox library \cite{kidger_equinox_2021} to define and train the embedded neural network within the hybrid (physics plus NN) model. Specifically, the objective of the demonstration is to estimate an initial density field such that, after $t_k$ timesteps of nonlinear fluid evolution (including intra and inter-phase interactions arising from Shan-Chen forces), the resulting density field approximates a droplet suspended in its vapor (henceforth referred to as the ground truth density). The simulation is performed on a $100 \times 100$ grid, employing the BGK collision model and the Van der Waals equation of state with parameters from section \ref{laplace_law}, at a reduced temperature of $T_r = 0.8$. The corresponding liquid and vapor densities are $\rho_l = 6.76447$ and $\rho_g = 0.838834$, respectively. The ground-truth density field ($\rho_{\mathrm{ground}}$) to be reproduced at $t = t_k$ is defined analogously to equation \ref{eq:laplace_init_density}:
\protect\begin{equation}\label{eq:ml_init_density}
    \begin{split}
        \rho_{\mathrm{ground}} = \frac{\rho_l - \rho_g}{2} - \frac{\rho_l - \rho_g}{2} \text{tanh}\Big(2 \frac{d -r}{w}\Big)
    \end{split}
\end{equation}
Here, $r = 25$ and $w = 4$. Periodic boundary conditions are applied in all directions, and the initial velocity field is set to zero. The initial density field required to reproduce this target or ground truth density distribution is obtained using a feedforward multi-layer perceptron with hidden layers of 512, 256, 256, and 512 neurons, respectively, as a representative example. Training is performed for 300 epochs using the adaptive moment estimation (ADAM) optimizer \cite{kingma_adam_2017} with a learning rate of $1 \times 10^{-3}$. For this optimization, we use the mean squared error (MSE) loss function, which is defined as:
\protect\begin{equation}\label{eq:mean_square_error}
    \mathcal{L} = \frac{1}{nx \cdot ny}\sum \left(\sum_{i = 1}^q f_i^{t_{k}} - \rho_{\mathrm{ground}}\right)^2
\end{equation}
For $t_k = 900$, the mean squared error during training is presented in Fig. \ref{fig:ml-error}, and the results obtained from solving the above optimization problem are shown in Fig. \ref{fig:consolidated}.

This demonstration illustrates how automatic differentiation \cite{jax2018github}, when combined with the LBM framework, can be employed to address inverse problems in multiphase flows. More physically realistic results could potentially be achieved by augmenting the loss function in equation \ref{eq:mean_square_error} with additional physically informed constraints such as Galilean and rotational invariance, mass and momentum conservation, among others \cite{corbetta_toward_2023}. The framework can be extended to more complex geometries, such as porous media, where it may be applied for topology optimization to design structures with optimal flow and transport properties \cite{kong_adjoint_2025}, or to establish correlations between pore-scale microstructure and flow behavior \cite{graczyk_predicting_2020}, among other applications.
\protect\begin{figure}[H]
\centering
\subfloat[\label{fig:ml-workflow}]{
    \noindent\includegraphics[width=8cm, height=2.03cm]{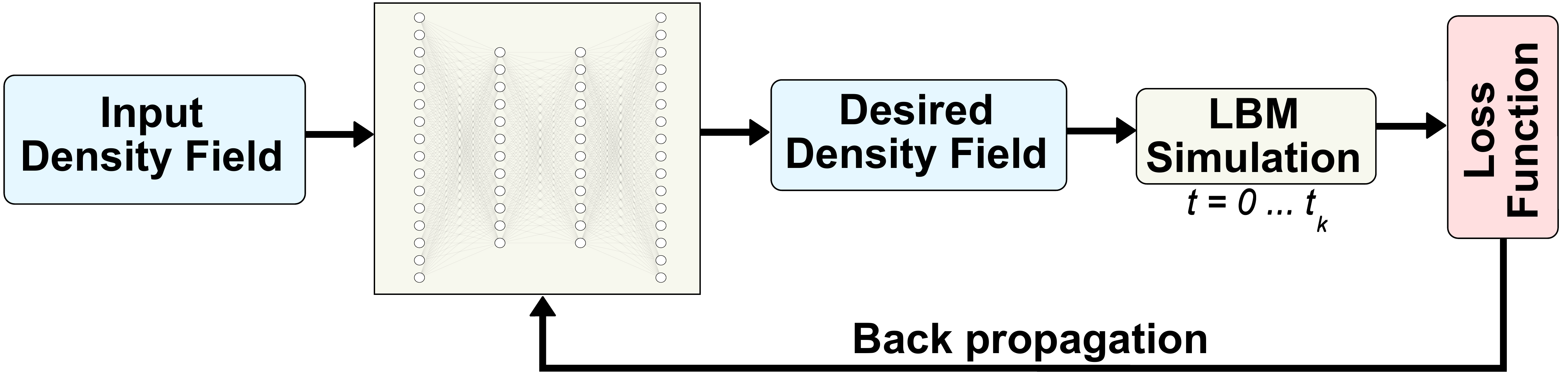}
}
\vspace{1em}
\centering
\subfloat[\label{fig:ml-error}]{
    \noindent\includegraphics[scale=0.3]{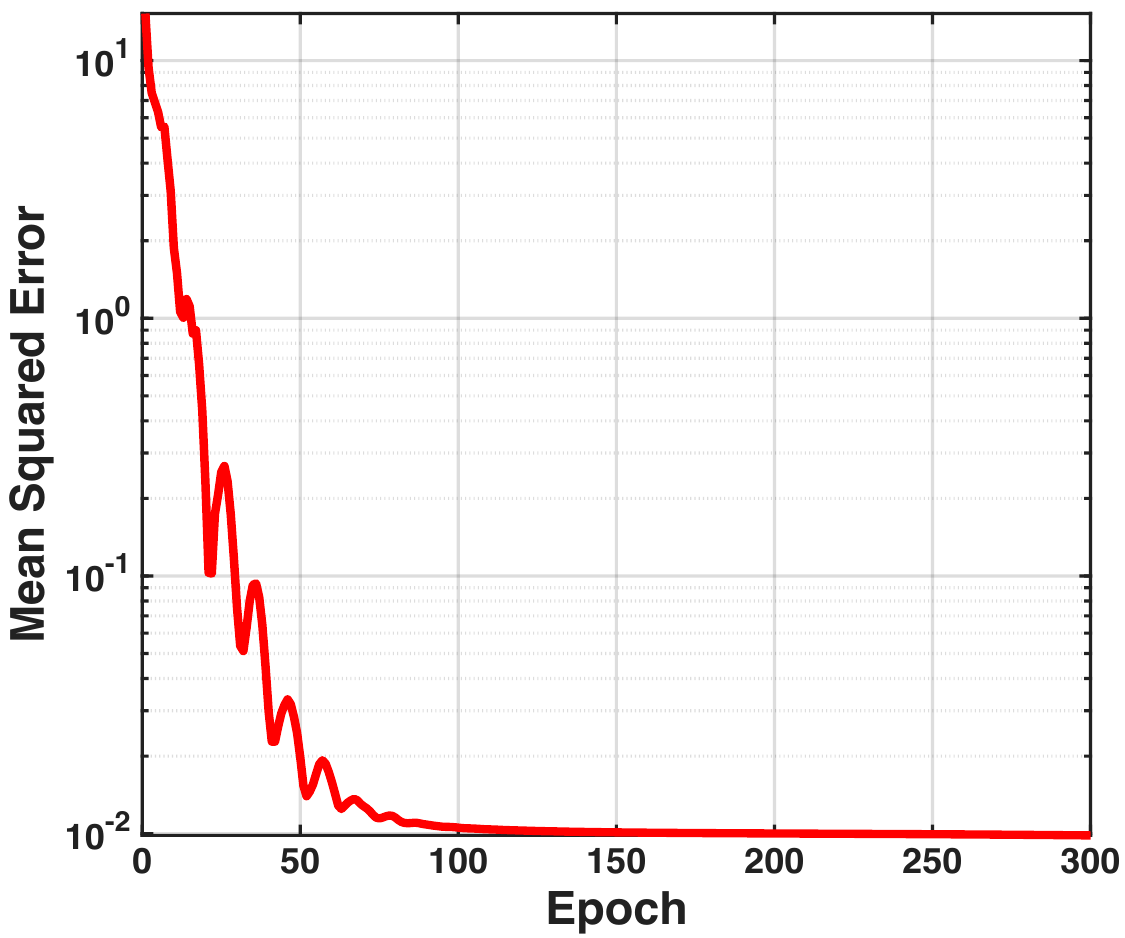}
}
\caption{\protect\add[PP]{(a) Diagram of the workflow used to train the neural-network. (b) Mean-squared error (MSE) decay with training epochs.}}
\label{fig:workflow-training}
\end{figure}}
\protect
\section{Performance Benchmarks}
The performance of JAX-LaB was evaluated across various single-GPU and multi-GPU systems. Given the library's heavy dependence on JAX, the latest stable release available at the time of writing (version 0.6.0) was used to take advantage of recent performance enhancements and improvements. Simulations are performed using the D3Q19 lattice and the BGK collision model, with all I/O operations excluded. Unless specified otherwise, computations are carried out in single precision (\verb|f32/f32|). The test case involves a droplet of radius 30 lattice units suspended in its vapor (Fig. \ref{fig:droplet_test}), using a pseudopotential defined as $\psi = \exp(-1/\rho)$, resulting in a liquid-vapor density ratio of approximately $\rho_l/\rho_v \approx 7.6$. Periodic boundary conditions are applied in all directions. This test case was selected for performance benchmarking because it involves a non-linear pseudopotential and can be easily scaled to match the GPU's computational capacity. The performance of the code is quantified using million lattice updates per second (MLUPS), which is defined as: 
\begin{equation}
    \text{MLUPS} = \frac{\text{No. of Components} \times nx \times ny \times nz}{\text{Compute Time(s)} \times 10^6}
\end{equation}
During computation of MLUPS, the first iteration is not considered to account for the time taken for the just-in-time (JIT) compilation of the main-loop \cite{ataei_xlb_2024}.
\subsection{Single GPU performance}
\begin{figure}[H]
    \centering
    \noindent\includegraphics[scale=0.5]{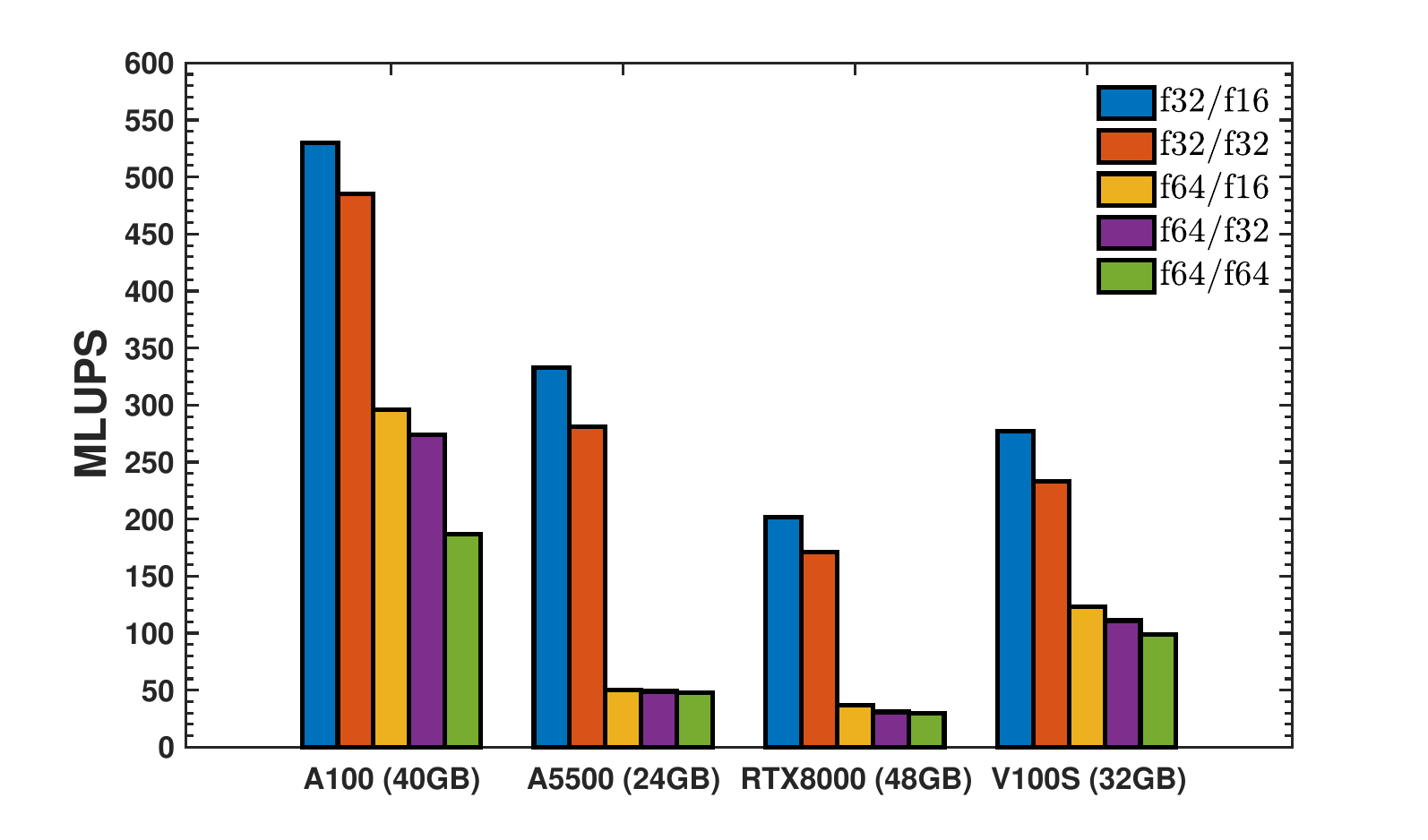}
    \caption{Single-GPU performance of JAX-LaB on different GPU architectures, across various mixed precision configurations (compute/write) for a droplet suspended in vapor.}
    \label{fig:single_GPU_perf}
\end{figure}

In this section, we evaluate the performance of JAX-LaB on single-GPU systems. Figure \ref{fig:single_GPU_perf} presents MLUPS measurements across a range of GPUs, from desktop to server-grade, under different memory configurations and mixed-precision modes. The domain sizes are chosen to fully utilize the computational throughput of each device, though they do not reflect the maximum problem sizes each GPU is capable of handling. Across all tested hardware, mixed-precision (f32/f16) maximizes throughput but sacrifices numerical fidelity, making it unsuitable for simulations with high density ratios, where vapor-phase densities may be inaccurately represented. Single-precision (f32/f32) offers the best compromise between speed and accuracy, even under extreme density contrasts. Among the GPUs examined, the A100 (Ampere architecture) delivers the highest MLUPS in both single- and double-precision modes. As anticipated, double-precision workloads result in a significant performance drop across all GPUs evaluated in this study.
\subsection{Multi-GPU scaling}
\begin{figure}
\centering
\subfloat[][]{
    \centering
    \includegraphics[width=6cm, height=5cm]{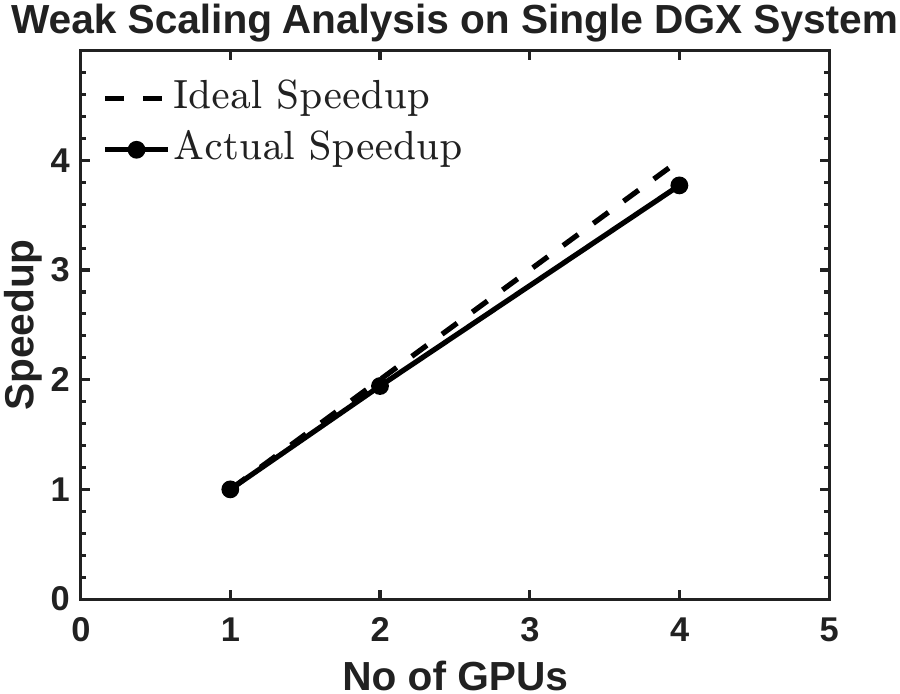}
    \label{fig:weak_scaling}
}
\subfloat[][]{
    \centering
    \includegraphics[width=6cm, height=5cm]{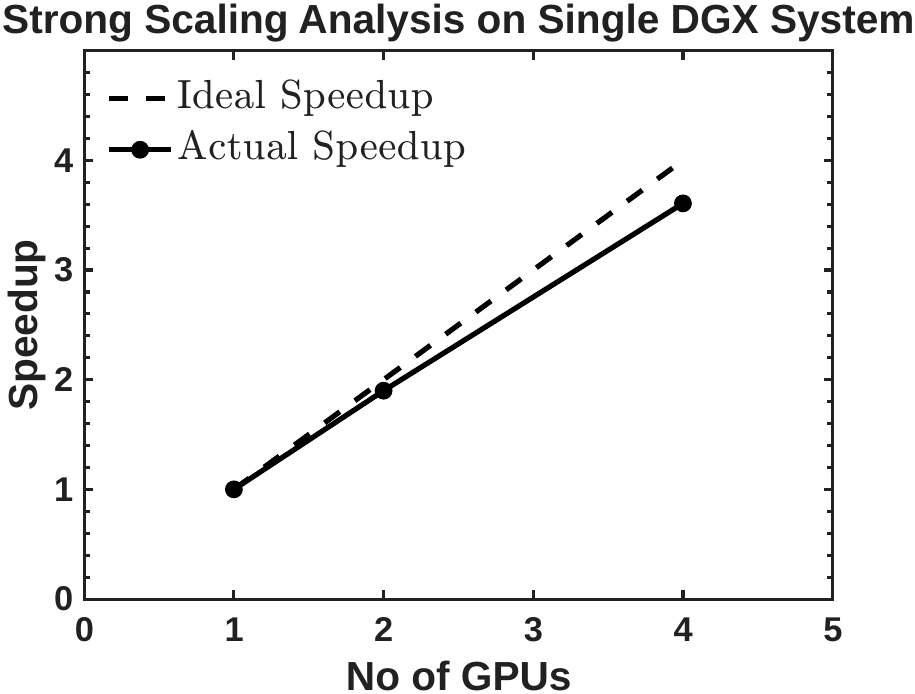}
    \label{fig:strong_scaling}
}
\caption{Scaling studies on A100 (40GB) GPUs: (a) Weak scaling with domain of sizes $416n \times 416^2$ where $n$ is the number of GPUs. (b) Strong scaling with cubic domain of size $320^3$.}
\label{fig:scaling_test}
\end{figure}

We evaluate the multi-GPU scaling performance of the JAX-LaB library on a DGX system equipped with A100 (40GB) GPUs. The test setup mirrors the previously described configuration, but operates in single precision (\verb|f32/f32|). Figure \ref{fig:weak_scaling} illustrates the results of the weak scaling test, where the domain size increases proportionally with the number of GPUs along the $x$-direction, following the form $416n \times 416^2$. The results exhibit excellent scaling behavior, maintaining over 94\% efficiency with 4 GPUs. Figure \ref{fig:strong_scaling} presents the strong scaling performance for a fixed domain size of $320^3$ as the GPU count increases. This test also shows high efficiency, achieving nearly 90\% with 4 GPUs. These tests reveal the capability of JAX-LaB to model very large multiphase systems (4 GPU weak scaling tests involved more than 287 million points).
\add[PP]{\protect\subsection{Distributed scaling}
In this section, we examine the distributed weak scaling performance of JAX-LaB on a cluster of DGX A100 (40 GB) GPU nodes, each equipped with two GPUs. The scaling test is conducted on up to four nodes, corresponding to a total of eight GPUs, using the same test configuration as in previous sections. All simulations are performed in single precision, with domain sizes scaled according to the number of available GPUs. To execute the code, individual JAX processes (initialized with jax.distributed.initialize) are launched on each node using the Message Passing Interface (MPI). MPI is the only configuration required for JAX-LaB to run on distributed systems. The results, presented in Fig. \ref{fig:dgx_scaling_test}, demonstrate scaling efficiency of 70\% (weak scaling) and 66\% (strong scaling) up to eight GPUs. The reduction in performance with an increasing number of nodes can be attributed to communication latency between nodes.
\begin{figure}
\centering
\subfloat[][]{
    \centering
    \includegraphics[width=7cm, height=5.8cm]{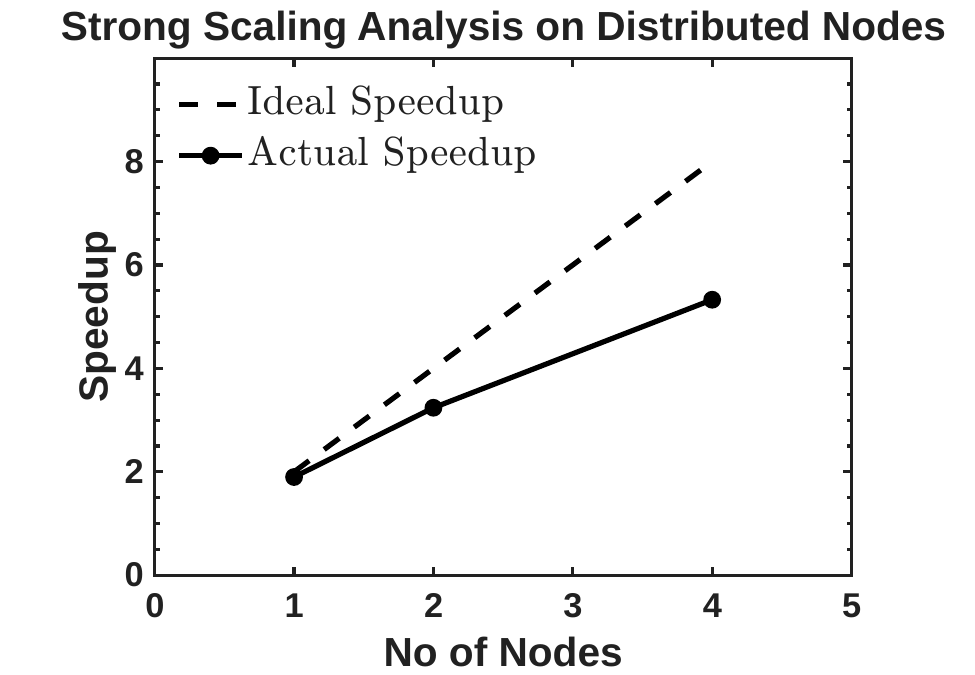}
    \label{fig:strong_scaling_dgx}
}
\subfloat[][]{
    \centering
    \includegraphics[width=7cm, height=5.8cm]{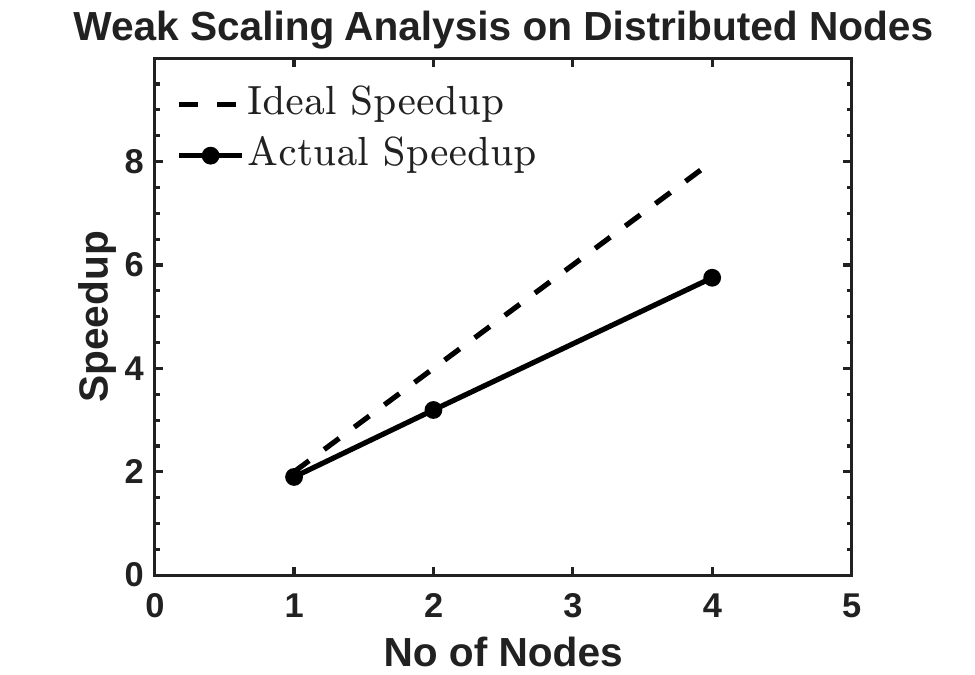}
    \label{fig:weak_scaling_dgx}
}
\caption{\protect\add[PP]{(a) Distributed scaling of JAX-LaB on a DGX A100 cluster (single precision): (a) Strong scaling with cubic domain of size $320^3$. (b) Weak scaling with domain of sizes $412n \times 412^2$ where $n$ is the number of nodes.}}
\label{fig:dgx_scaling_test}
\end{figure}}
\section{Conclusions}
This paper introduces JAX-LaB, a JAX-based, multi-GPU LBM framework that extends XLB by incorporating support for multiphase and multicomponent systems, 
using the well-established Shan-Chen pseudopotential method. However, rather than relying on the original Shan-Chen formulation, JAX-LaB implements a revised version that integrates a suite of improvements proposed in the literature. These enhancements include: (i) direct incorporation of the equation of state, (ii) an accurate force scheme capable of handling large density ratios while suppressing spurious currents, (iii) a pressure-tensor modification technique that enables independent control of surface tension without altering the density ratio (a key limitation of the original model), and (iv) an improved wetting boundary condition that supports a wide range of contact angles with minimal spurious velocities. These features are particularly valuable for simulating hydrological and geological flows, which often involve multiphase mixtures with diverse wettability and large liquid-vapor density ratios; for example, permeability, drainage, and imbibition simulations for sandstone and sphere pack porous media are showcased in this paper. Inheriting the modular and extensible design of XLB, JAX-LaB supports an arbitrary number of components by employing a list-based data structure (\textit{pytrees}). The codebase has been validated against several analytical and numerical benchmarks across a variety of domains, demonstrating strong agreement in all cases. \add[PP]{We demonstrated the differentiability of the code by utilizing the automatic differentiation capabilities of JAX to integrate machine learning with a physics-based code, employing this unified workflow to achieve fluid flow control through deep learning.} These test cases are included as reproducible examples within the repository to facilitate user adoption and verification. Furthermore, the scalability of the codebase on massively parallel hardware such as GPUs has been rigorously benchmarked. These tests evaluate both single- and double-precision performance on various GPU architectures and characterize multi-GPU scalability of the codebase using weak and strong scaling tests for single and distributed GPU nodes.

JAX-LaB is actively maintained and distributed under the Apache License at \url{https://github.com/piyush-ppradhan/JAX-LaB}. Planned features include sparse array support to reduce memory overhead, implementation of thermal LBM for non-isothermal flows, and integration of turbulence modeling via Large Eddy Simulation (LES). These developments aim to extend JAX-LaB's applicability to a broad range of fluid flow modeling problems relevant to geoscience and engineering applications.
\section*{Conflict of Interest}
The authors declare no conflicts of interest relevant to this study.
\section*{Data Availability Statement} 
JAX-LaB is released under the Apache License and is available at \url{https://github.com/piyush-ppradhan/JAX-LaB}. Since the library is under active development, the exact version of the code used to produce the results in this paper is made available separately at \cite{jax_lab_zenodo}. This release includes the relevant JAX-LaB source code, example scripts, and visualization figures used in the various showcase cases.

\acknowledgments
The single-GPU performance and multi-GPU scaling tests are performed on the Ginsburg cluster at Columbia University. The conceptualization and development of JAX-LaB by P.P., P.G., and S.K. was supported in part by the European Research Council (ERC) Synergy Grant "Understanding and modeling the Earth System with Machine Learning (USMILE)" under the Horizon 2020 research and innovation programme under Grant Agreement 855187 (P.G.) and the U.S. Department of Energy, Office of Science, Office of Basic Energy Sciences, Geosciences program under Award Number DE‐SC0025348 (S.K.). P.G. acknowledges additional funding from the National Science Foundation Science and Technology Center, Learning the Earth with Artificial Intelligence and Physics, LEAP (Grant 2019625). S.K. acknowledges additional funding from the American Chemical Society Petroleum Research Fund Grant 66777-DNI9 and the Columbia University Qiu Zhong Wei Research Project fund.

%
%


%
%
%
%
%

\bibliography{references}

@misc{jax2018github,
  author = {James Bradbury and Roy Frostig and Peter Hawkins and Matthew James Johnson and Chris Leary and Dougal Maclaurin and George Necula and Adam Paszke and Jake Vander{P}las and Skye Wanderman-{M}ilne and Qiao Zhang},
  title = {{JAX}: composable transformations of {P}ython+{N}um{P}y programs},
  url = {http://github.com/jax-ml/jax},
  version = {0.3.13},
  year = {2018},
}

@misc{jax_lab_zenodo,
    author = {Piyush Pradhan and Pierre Gentine and Shaina Kelly},
    title = {{JAX}-{L}a{B}: A {P}ython-based, {A}ccelerated, {D}ifferentiable {M}assively {P}arallel {L}attice {B}oltzmann {L}ibrary for {M}odeling {M}ultiphase and {M}ultiphysics {F}lows \& {P}hysics-{B}ased {M}achine {L}earning},
    url = {https://doi.org/10.5281/zenodo.17613708},
    year = {2025}
}

@misc{phantomgaze,
  author = {O. Hennigh},
  title = {PhantomGaze: a GPU-accelerated rendering engine for scientific visualization},
  url = {https://github.com/loliverhennigh/PhantomGaze},
  year = {2023}
}

@misc{drp_geometry,
  author = {E. Santos, Javier and Chang, Bernard  and Kang, Qinjun and Viswanathan, Hari and Lubbers, Nicholas and Gigliotti, Alex and Prodanovic, Masa},
  title = {3D Dataset of Simulations},
  url = {https://www.doi.org/10.17612/93pd-y471},
  year = {2025}
}

@article{harris_array_2020,
    title = {Array programming with {NumPy}},
    volume = {585},
    copyright = {2020 The Author(s)},
    issn = {1476-4687},
    url = {https://www.nature.com/articles/s41586-020-2649-2},
    doi = {10.1038/s41586-020-2649-2},
    language = {en},
    number = {7825},
    journal = {Nature},
    author = {Harris, Charles R. and Millman, K. Jarrod and van der Walt, Stéfan J. and Gommers, Ralf and Virtanen, Pauli and Cournapeau, David and Wieser, Eric and Taylor, Julian and Berg, Sebastian and Smith, Nathaniel J. and Kern, Robert and Picus, Matti and Hoyer, Stephan and van Kerkwijk, Marten H. and Brett, Matthew and Haldane, Allan and del Río, Jaime Fernández and Wiebe, Mark and Peterson, Pearu and Gérard-Marchant, Pierre and Sheppard, Kevin and Reddy, Tyler and Weckesser, Warren and Abbasi, Hameer and Gohlke, Christoph and Oliphant, Travis E.},
    month = sep,
    year = {2020},
    note = {Publisher: Nature Publishing Group},
    pages = {357--362},
}

@misc{paszke_pytorch_2019,
    title = {{PyTorch}: {An} {Imperative} {Style}, {High}-{Performance} {Deep} {Learning} {Library}},
    shorttitle = {{PyTorch}},
    url = {http://arxiv.org/abs/1912.01703},
    doi = {10.48550/arXiv.1912.01703},
    publisher = {arXiv},
    author = {Paszke, Adam and Gross, Sam and Massa, Francisco and Lerer, Adam and Bradbury, James and Chanan, Gregory and Killeen, Trevor and Lin, Zeming and Gimelshein, Natalia and Antiga, Luca and Desmaison, Alban and Köpf, Andreas and Yang, Edward and DeVito, Zach and Raison, Martin and Tejani, Alykhan and Chilamkurthy, Sasank and Steiner, Benoit and Fang, Lu and Bai, Junjie and Chintala, Soumith},
    month = dec,
    year = {2019},
    note = {arXiv:1912.01703 [cs, stat]},
}

@misc{abadi_tensorflow_2016,
    title = {{TensorFlow}: {A} system for large-scale machine learning},
    shorttitle = {{TensorFlow}},
    url = {http://arxiv.org/abs/1605.08695},
    doi = {10.48550/arXiv.1605.08695},
    publisher = {arXiv},
    author = {Abadi, Martín and Barham, Paul and Chen, Jianmin and Chen, Zhifeng and Davis, Andy and Dean, Jeffrey and Devin, Matthieu and Ghemawat, Sanjay and Irving, Geoffrey and Isard, Michael and Kudlur, Manjunath and Levenberg, Josh and Monga, Rajat and Moore, Sherry and Murray, Derek G. and Steiner, Benoit and Tucker, Paul and Vasudevan, Vijay and Warden, Pete and Wicke, Martin and Yu, Yuan and Zheng, Xiaoqiang},
    month = may,
    year = {2016},
    note = {arXiv:1605.08695 [cs]},
}

@article{mittal_immersed_2005,
    title = {Immersed {Boundary} {Methods}},
    volume = {37},
    issn = {0066-4189, 1545-4479},
    url = {https://www.annualreviews.org/content/journals/10.1146/annurev.fluid.37.061903.175743},
    doi = {10.1146/annurev.fluid.37.061903.175743},
    language = {en},
    number = {Volume 37, 2005},
    journal = {Annual Review of Fluid Mechanics},
    author = {Mittal, Rajat and Iaccarino, Gianluca},
    month = jan,
    year = {2005},
    note = {Publisher: Annual Reviews},
    pages = {239--261},
}

@article{zhang_lattice_2011,
    title = {Lattice {Boltzmann} method for microfluidics: models and applications},
    volume = {10},
    issn = {1613-4990},
    shorttitle = {Lattice {Boltzmann} method for microfluidics},
    url = {https://doi.org/10.1007/s10404-010-0624-1},
    doi = {10.1007/s10404-010-0624-1},
    language = {en},
    number = {1},
    journal = {Microfluidics and Nanofluidics},
    author = {Zhang, Junfeng},
    month = jan,
    year = {2011},
    pages = {1--28},
}

@article{worner_numerical_2012,
    title = {Numerical modeling of multiphase flows in microfluidics and micro process engineering: a review of methods and applications},
    volume = {12},
    issn = {1613-4990},
    shorttitle = {Numerical modeling of multiphase flows in microfluidics and micro process engineering},
    url = {https://doi.org/10.1007/s10404-012-0940-8},
    doi = {10.1007/s10404-012-0940-8},
    language = {en},
    number = {6},
    journal = {Microfluidics and Nanofluidics},
    author = {Wörner, Martin},
    month = may,
    year = {2012},
    pages = {841--886},
}

@article{aidun_lattice-boltzmann_2010,
    title = {Lattice-{Boltzmann} {Method} for {Complex} {Flows}},
    volume = {42},
    issn = {0066-4189, 1545-4479},
    url = {https://www.annualreviews.org/content/journals/10.1146/annurev-fluid-121108-145519},
    doi = {10.1146/annurev-fluid-121108-145519},
    language = {en},
    number = {Volume 42, 2010},
    journal = {Annual Review of Fluid Mechanics},
    author = {Aidun, Cyrus K. and Clausen, Jonathan R.},
    month = jan,
    year = {2010},
    note = {Publisher: Annual Reviews},
    pages = {439--472},
}

@article{petersen_lattice_2021,
    title = {On the lattice {Boltzmann} method and its application to turbulent, multiphase flows of various fluids including cryogens: {A} review},
    volume = {33},
    issn = {1070-6631},
    shorttitle = {On the lattice {Boltzmann} method and its application to turbulent, multiphase flows of various fluids including cryogens},
    url = {https://doi.org/10.1063/5.0046938},
    doi = {10.1063/5.0046938},
    number = {4},
    journal = {Physics of Fluids},
    author = {Petersen, K. J. and Brinkerhoff, J. R.},
    month = apr,
    year = {2021},
    pages = {041302},
}

@article{feng_hybrid_2019,
    title = {Hybrid recursive regularized lattice {Boltzmann} simulation of humid air with application to meteorological flows},
    volume = {100},
    url = {https://link.aps.org/doi/10.1103/PhysRevE.100.023304},
    doi = {10.1103/PhysRevE.100.023304},
    number = {2},
    journal = {Physical Review E},
    author = {Feng, Yongliang and Boivin, Pierre and Jacob, Jérôme and Sagaut, Pierre},
    month = aug,
    year = {2019},
    note = {Publisher: American Physical Society},
    pages = {023304},
}

@article{coon_taxila_2014,
    title = {Taxila {LBM}: a parallel, modular lattice {Boltzmann} framework for simulating pore-scale flow in porous media},
    volume = {18},
    issn = {1573-1499},
    shorttitle = {Taxila {LBM}},
    url = {https://doi.org/10.1007/s10596-013-9379-6},
    doi = {10.1007/s10596-013-9379-6},
    language = {en},
    number = {1},
    journal = {Computational Geosciences},
    author = {Coon, Ethan T. and Porter, Mark L. and Kang, Qinjun},
    month = feb,
    year = {2014},
    pages = {17--27},
}

@article{latt_palabos_2021,
    series = {Development and {Application} of {Open}-source {Software} for {Problems} with {Numerical} {PDEs}},
    title = {Palabos: {Parallel} {Lattice} {Boltzmann} {Solver}},
    volume = {81},
    issn = {0898-1221},
    shorttitle = {Palabos},
    doi = {10.1016/j.camwa.2020.03.022},
    journal = {Computers \& Mathematics with Applications},
    author = {Latt, Jonas and Malaspinas, Orestis and Kontaxakis, Dimitrios and Parmigiani, Andrea and Lagrava, Daniel and Brogi, Federico and Belgacem, Mohamed Ben and Thorimbert, Yann and Leclaire, Sébastien and Li, Sha and Marson, Francesco and Lemus, Jonathan and Kotsalos, Christos and Conradin, Raphaël and Coreixas, Christophe and Petkantchin, Rémy and Raynaud, Franck and Beny, Joël and Chopard, Bastien},
    month = jan,
    year = {2021},
    pages = {334--350},
}

@article{krause_openlbopen_2021,
    series = {Development and {Application} of {Open}-source {Software} for {Problems} with {Numerical} {PDEs}},
    title = {{OpenLB}—{Open} source lattice {Boltzmann} code},
    volume = {81},
    issn = {0898-1221},
    url = {https://www.sciencedirect.com/science/article/pii/S0898122120301875},
    doi = {10.1016/j.camwa.2020.04.033},
    journal = {Computers \& Mathematics with Applications},
    author = {Krause, Mathias J. and Kummerländer, Adrian and Avis, Samuel J. and Kusumaatmaja, Halim and Dapelo, Davide and Klemens, Fabian and Gaedtke, Maximilian and Hafen, Nicolas and Mink, Albert and Trunk, Robin and Marquardt, Jan E. and Maier, Marie-Luise and Haussmann, Marc and Simonis, Stephan},
    month = jan,
    year = {2021},
    pages = {258--288},
}

@article{chen_inertial_2019,
    title = {Inertial {Effects} {During} the {Process} of {Supercritical} {CO2} {Displacing} {Brine} in a {Sandstone}: {Lattice} {Boltzmann} {Simulations} {Based} on the {Continuum}-{Surface}-{Force} and {Geometrical} {Wetting} {Models}},
    volume = {55},
    copyright = {©2019. American Geophysical Union. All Rights Reserved.},
    issn = {1944-7973},
    shorttitle = {Inertial {Effects} {During} the {Process} of {Supercritical} {CO2} {Displacing} {Brine} in a {Sandstone}},
    url = {https://onlinelibrary.wiley.com/doi/abs/10.1029/2019WR025746},
    doi = {10.1029/2019WR025746},
    language = {en},
    number = {12},
    journal = {Water Resources Research},
    author = {Chen, Yu and Valocchi, Albert J. and Kang, Qinjun and Viswanathan, Hari S.},
    year = {2019},
    pages = {11144--11165},
}

@article{januszewski_sailfish_2014,
    title = {Sailfish: {A} flexible multi-{GPU} implementation of the lattice {Boltzmann} method},
    volume = {185},
    issn = {0010-4655},
    shorttitle = {Sailfish},
    url = {https://www.sciencedirect.com/science/article/pii/S0010465514001520},
    doi = {10.1016/j.cpc.2014.04.018},
    number = {9},
    journal = {Computer Physics Communications},
    author = {Januszewski, M. and Kostur, M.},
    month = sep,
    year = {2014},
    pages = {2350--2368},
}

@article{bauer_walberla_2021,
    series = {Development and {Application} of {Open}-source {Software} for {Problems} with {Numerical} {PDEs}},
    title = {{waLBerla}: {A} block-structured high-performance framework for multiphysics simulations},
    volume = {81},
    issn = {0898-1221},
    shorttitle = {{waLBerla}},
    url = {https://www.sciencedirect.com/science/article/pii/S0898122120300146},
    doi = {10.1016/j.camwa.2020.01.007},
    journal = {Computers \& Mathematics with Applications},
    author = {Bauer, Martin and Eibl, Sebastian and Godenschwager, Christian and Kohl, Nils and Kuron, Michael and Rettinger, Christoph and Schornbaum, Florian and Schwarzmeier, Christoph and Thönnes, Dominik and Köstler, Harald and Rüde, Ulrich},
    month = jan,
    year = {2021},
    pages = {478--501},
}

@article{santos_mplbm-ut_2022,
    title = {{MPLBM}-{UT}: {Multiphase} {LBM} library for permeable media analysis},
    volume = {18},
    issn = {2352-7110},
    shorttitle = {{MPLBM}-{UT}},
    url = {https://www.sciencedirect.com/science/article/pii/S2352711022000668},
    doi = {10.1016/j.softx.2022.101097},
    journal = {SoftwareX},
    author = {Santos, Javier E. and Gigliotti, Alex and Bihani, Abhishek and Landry, Christopher and Hesse, Marc A. and Pyrcz, Michael J. and Prodanović, Maša},
    month = jun,
    year = {2022},
    pages = {101097},
}

@article{calzavarini_eulerianlagrangian_2019,
    title = {Eulerian–{Lagrangian} fluid dynamics platform: {The} ch4-project},
    volume = {1},
    issn = {2665-9638},
    shorttitle = {Eulerian–{Lagrangian} fluid dynamics platform},
    url = {https://www.sciencedirect.com/science/article/pii/S2665963819300028},
    doi = {10.1016/j.simpa.2019.100002},
    journal = {Software Impacts},
    author = {Calzavarini, Enrico},
    month = sep,
    year = {2019},
    pages = {100002},
}

@inproceedings{luebke_cuda_2008,
    title = {{CUDA}: {Scalable} parallel programming for high-performance scientific computing},
    shorttitle = {{CUDA}},
    url = {https://ieeexplore.ieee.org/document/4541126},
    doi = {10.1109/ISBI.2008.4541126},
    booktitle = {2008 5th {IEEE} {International} {Symposium} on {Biomedical} {Imaging}: {From} {Nano} to {Macro}},
    author = {Luebke, David},
    month = may,
    year = {2008},
    note = {ISSN: 1945-8452},
    pages = {836--838},
}

@phdthesis{lehmann_high_2019,
    title = {High {Performance} {Free} {Surface} {LBM} on {GPUs}},
    author = {Lehmann, Moritz},
    month = dec,
    year = {2019},
    doi = {10.15495/EPub_UBT_00005400},
}

@article{stone_opencl_2010,
    title = {{OpenCL}: {A} {Parallel} {Programming} {Standard} for {Heterogeneous} {Computing} {Systems}},
    volume = {12},
    issn = {1558-366X},
    shorttitle = {{OpenCL}},
    url = {https://ieeexplore.ieee.org/document/5457293},
    doi = {10.1109/MCSE.2010.69},
    number = {3},
    journal = {Computing in Science \& Engineering},
    author = {Stone, John E. and Gohara, David and Shi, Guochun},
    month = may,
    year = {2010},
    note = {Conference Name: Computing in Science \& Engineering},
    pages = {66--73},
}

@inproceedings{bedrunka_lettuce_2021,
    address = {Cham},
    title = {Lettuce: {PyTorch}-{Based} {Lattice} {Boltzmann} {Framework}},
    isbn = {978-3-030-90539-2},
    shorttitle = {Lettuce},
    doi = {10.1007/978-3-030-90539-2_3},
    language = {en},
    booktitle = {High {Performance} {Computing}},
    publisher = {Springer International Publishing},
    author = {Bedrunka, Mario Christopher and Wilde, Dominik and Kliemank, Martin and Reith, Dirk and Foysi, Holger and Krämer, Andreas},
    editor = {Jagode, Heike and Anzt, Hartwig and Ltaief, Hatem and Luszczek, Piotr},
    year = {2021},
    pages = {40--55},
}

@article{yang_taichi-lbm3d_2022,
    title = {Taichi-{LBM3D}: {A} {Single}-{Phase} and {Multiphase} {Lattice} {Boltzmann} {Solver} on {Cross}-{Platform} {Multicore} {CPU}/{GPUs}},
    shorttitle = {Taichi-{LBM3D}},
    doi = {10.3390/fluids7080270},
    journal = {Fluids},
    author = {Yang, Jianhui and Xu, Yi and Yang, Liang},
    month = aug,
    year = {2022},
}

@article{bezgin_jax-fluids_2023,
    title = {{JAX}-{Fluids}: {A} fully-differentiable high-order computational fluid dynamics solver for compressible two-phase flows},
    volume = {282},
    issn = {0010-4655},
    shorttitle = {{JAX}-{Fluids}},
    url = {https://www.sciencedirect.com/science/article/pii/S0010465522002466},
    doi = {10.1016/j.cpc.2022.108527},
    journal = {Computer Physics Communications},
    author = {Bezgin, Deniz A. and Buhendwa, Aaron B. and Adams, Nikolaus A.},
    month = jan,
    year = {2023},
    pages = {108527},
}

@article{kupershtokh_equations_2009,
    title = {On equations of state in a lattice {Boltzmann} method},
    volume = {58},
    copyright = {https://www.elsevier.com/tdm/userlicense/1.0/},
    issn = {08981221},
    url = {https://linkinghub.elsevier.com/retrieve/pii/S0898122109001011},
    doi = {10.1016/j.camwa.2009.02.024},
    language = {en},
    number = {5},
    journal = {Computers \& Mathematics with Applications},
    author = {Kupershtokh, A.L. and Medvedev, D.A. and Karpov, D.I.},
    month = sep,
    year = {2009},
    pages = {965--974},
}

@article{shan_lattice_1993,
    title = {Lattice {Boltzmann} model for simulating flows with multiple phases and components},
    volume = {47},
    url = {https://link.aps.org/doi/10.1103/PhysRevE.47.1815},
    doi = {10.1103/PhysRevE.47.1815},
    number = {3},
    journal = {Physical Review E},
    author = {Shan, Xiaowen and Chen, Hudong},
    month = mar,
    year = {1993},
    note = {Publisher: American Physical Society},
    pages = {1815--1819},
}

@article{gong_numerical_2012,
    title = {Numerical investigation of droplet motion and coalescence by an improved lattice {Boltzmann} model for phase transitions and multiphase flows},
    volume = {53},
    issn = {0045-7930},
    url = {https://www.sciencedirect.com/science/article/pii/S0045793011002994},
    doi = {10.1016/j.compfluid.2011.09.013},
    journal = {Computers \& Fluids},
    author = {Gong, Shuai and Cheng, Ping},
    month = jan,
    year = {2012},
    pages = {93--104},
}

@article{zhang_lattice_2003,
    title = {Lattice {Boltzmann} method for simulations of liquid-vapor thermal flows},
    volume = {67},
    url = {https://link.aps.org/doi/10.1103/PhysRevE.67.066711},
    doi = {10.1103/PhysRevE.67.066711},
    number = {6},
    journal = {Physical Review E},
    author = {Zhang, Raoyang and Chen, Hudong},
    month = jun,
    year = {2003},
    note = {Publisher: American Physical Society},
    pages = {066711},
}

@misc{hu_surface_2014,
    title = {Surface {Tension} {Adjustment} in a {Pseudo}-{Potential} {Lattice} {Boltzmann} {Model}},
    url = {http://arxiv.org/abs/1412.7228},
    doi = {10.48550/arXiv.1412.7228},
    publisher = {arXiv},
    author = {Hu, Anjie and Li, Longjian and Uddin, Rizwan},
    month = dec,
    year = {2014},
    note = {arXiv:1412.7228},
}

@article{qin_effective_2022,
    title = {An effective pseudo-potential lattice {Boltzmann} model with extremely large density ratio and adjustable surface tension},
    volume = {34},
    issn = {1070-6631},
    url = {https://doi.org/10.1063/5.0123727},
    doi = {10.1063/5.0123727},
    number = {11},
    journal = {Physics of Fluids},
    author = {Qin, Zhangrong and Zhu, Jianfei and Chen, Wenbo and Li, Chengsheng and Wen, Binghai},
    month = nov,
    year = {2022},
    pages = {113328},
}

@article{yuan_equations_2006,
    title = {Equations of state in a lattice {Boltzmann} model},
    volume = {18},
    issn = {1070-6631},
    url = {https://doi.org/10.1063/1.2187070},
    doi = {10.1063/1.2187070},
    number = {4},
    journal = {Physics of Fluids},
    author = {Yuan, Peng and Schaefer, Laura},
    month = apr,
    year = {2006},
    pages = {042101},
}

@article{sbragaglia_generalized_2007,
    title = {Generalized lattice {Boltzmann} method with multirange pseudopotential},
    volume = {75},
    url = {https://link.aps.org/doi/10.1103/PhysRevE.75.026702},
    doi = {10.1103/PhysRevE.75.026702},
    number = {2},
    journal = {Physical Review E},
    author = {Sbragaglia, M. and Benzi, R. and Biferale, L. and Succi, S. and Sugiyama, K. and Toschi, F.},
    month = feb,
    year = {2007},
    note = {Publisher: American Physical Society},
    pages = {026702},
}

@article{li_achieving_2013,
    title = {Achieving tunable surface tension in the pseudopotential lattice {Boltzmann} modeling of multiphase flows},
    volume = {88},
    url = {https://link.aps.org/doi/10.1103/PhysRevE.88.053307},
    doi = {10.1103/PhysRevE.88.053307},
    number = {5},
    journal = {Physical Review E},
    author = {Li, Qing and Luo, K. H.},
    month = nov,
    year = {2013},
    note = {Publisher: American Physical Society},
    pages = {053307},
}

@article{shan_pressure_2008,
    title = {Pressure tensor calculation in a class of nonideal gas lattice {Boltzmann} models},
    volume = {77},
    url = {https://link.aps.org/doi/10.1103/PhysRevE.77.066702},
    doi = {10.1103/PhysRevE.77.066702},
    number = {6},
    journal = {Physical Review E},
    author = {Shan, Xiaowen},
    month = jun,
    year = {2008},
    note = {Publisher: American Physical Society},
    pages = {066702},
}

@article{zhu_investigation_2024,
    title = {Investigation on enhanced density ratio recovery and numerical stability in real physical field under multi-component multiphase {LBM}},
    volume = {156},
    issn = {0735-1933},
    url = {https://www.sciencedirect.com/science/article/pii/S0735193324004354},
    doi = {10.1016/j.icheatmasstransfer.2024.107673},
    journal = {International Communications in Heat and Mass Transfer},
    author = {Zhu, Junhao and Dai, Zheng and Wang, Zhongyi and Chu, Shuguang and Wang, Meng},
    month = aug,
    year = {2024},
    pages = {107673},
}

@article{li_implementation_2019,
    title = {Implementation of contact angles in pseudopotential lattice {Boltzmann} simulations with curved boundaries},
    volume = {100},
    issn = {2470-0045, 2470-0053},
    url = {https://link.aps.org/doi/10.1103/PhysRevE.100.053313},
    doi = {10.1103/PhysRevE.100.053313},
    language = {en},
    number = {5},
    journal = {Physical Review E},
    author = {Li, Q. and Yu, Y. and Luo, Kai H.},
    month = nov,
    year = {2019},
    pages = {053313},
}

@article{hu_contact_2016,
    title = {Contact angle adjustment in equation-of-state-based pseudopotential model},
    volume = {93},
    url = {https://link.aps.org/doi/10.1103/PhysRevE.93.053307},
    doi = {10.1103/PhysRevE.93.053307},
    number = {5},
    journal = {Physical Review E},
    author = {Hu, Anjie and Li, Longjian and Uddin, Rizwan and Liu, Dong},
    month = may,
    year = {2016},
    note = {Publisher: American Physical Society},
    pages = {053307},
}

@article{ding_wetting_2007,
    title = {Wetting condition in diffuse interface simulations of contact line motion},
    volume = {75},
    url = {https://link.aps.org/doi/10.1103/PhysRevE.75.046708},
    doi = {10.1103/PhysRevE.75.046708},
    number = {4},
    journal = {Physical Review E},
    author = {Ding, Hang and Spelt, Peter D. M.},
    month = apr,
    year = {2007},
    note = {Publisher: American Physical Society},
    pages = {046708},
}

@misc{ataei_xlb_2023,
    title = {{XLB}: {Distributed} {Multi}-{GPU} {Lattice} {Boltzmann} {Simulation} {Framework} for {Differentiable} {Scientific} {Machine} {Learning}},
    shorttitle = {{XLB}},
    url = {http://arxiv.org/abs/2311.16080},
    doi = {10.48550/arXiv.2311.16080},
    publisher = {arXiv},
    author = {Ataei, Mohammadmehdi and Salehipour, Hesam},
    month = nov,
    year = {2023},
    note = {arXiv:2311.16080 [physics]},
}

@article{tang_multi-component_2024,
    title = {Multi-component multiphase lattice {Boltzmann} modeling of water purging during supercritical carbon dioxide extraction from geothermal reservoir pores},
    volume = {220},
    issn = {09601481},
    url = {https://linkinghub.elsevier.com/retrieve/pii/S0960148123015756},
    doi = {10.1016/j.renene.2023.119660},
    language = {en},
    journal = {Renewable Energy},
    author = {Tang, Youfei and Qiao, Zongliang and Cao, Yue and Si, Fengqi and Zhang, Chengbin},
    month = jan,
    year = {2024},
    pages = {119660},
}

@article{santos_3d_2022,
    title = {{3D} {Dataset} of binary images: {A} collection of synthetically created digital rock images of complex media},
    volume = {40},
    issn = {2352-3409},
    shorttitle = {{3D} {Dataset} of binary images},
    url = {https://www.sciencedirect.com/science/article/pii/S2352340922000099},
    doi = {10.1016/j.dib.2022.107797},
    journal = {Data in Brief},
    author = {Santos, Javier E. and Pyrcz, Michael J. and Prodanović, Maša},
    month = feb,
    year = {2022},
    pages = {107797},
}

@article{porter_multicomponent_2012,
    title = {Multicomponent interparticle-potential lattice {Boltzmann} model for fluids with large viscosity ratios},
    volume = {86},
    copyright = {http://link.aps.org/licenses/aps-default-license},
    issn = {1539-3755, 1550-2376},
    url = {https://link.aps.org/doi/10.1103/PhysRevE.86.036701},
    doi = {10.1103/PhysRevE.86.036701},
    language = {en},
    number = {3},
    journal = {Physical Review E},
    author = {Porter, Mark L. and Coon, E. T. and Kang, Q. and Moulton, J. D. and Carey, J. W.},
    month = sep,
    year = {2012},
    pages = {036701},
}

@article{schindelin_fiji_2012,
    title = {Fiji: an open-source platform for biological-image analysis},
    volume = {9},
    copyright = {2012 Springer Nature America, Inc.},
    issn = {1548-7105},
    shorttitle = {Fiji},
    url = {https://www.nature.com/articles/nmeth.2019},
    doi = {10.1038/nmeth.2019},
    language = {en},
    number = {7},
    journal = {Nature Methods},
    author = {Schindelin, Johannes and Arganda-Carreras, Ignacio and Frise, Erwin and Kaynig, Verena and Longair, Mark and Pietzsch, Tobias and Preibisch, Stephan and Rueden, Curtis and Saalfeld, Stephan and Schmid, Benjamin and Tinevez, Jean-Yves and White, Daniel James and Hartenstein, Volker and Eliceiri, Kevin and Tomancak, Pavel and Cardona, Albert},
    month = jul,
    year = {2012},
    note = {Publisher: Nature Publishing Group},
    pages = {676--682},
}

@article{halpern_boundary_1994,
    title = {Boundary {Element} {Analysis} of the {Time}-{Dependent} {Motion} of a {Semi}-infinite {Bubble} in a {Channel}},
    volume = {115},
    issn = {0021-9991},
    url = {https://www.sciencedirect.com/science/article/pii/S0021999184712022},
    doi = {10.1006/jcph.1994.1202},
    number = {2},
    journal = {Journal of Computational Physics},
    author = {Halpern, D. and Gaver, D. P.},
    month = dec,
    year = {1994},
    pages = {366--375},
}

@article{kang_immiscible_2004,
    title = {Immiscible displacement in a channel: simulations of fingering in two dimensions},
    volume = {27},
    issn = {0309-1708},
    shorttitle = {Immiscible displacement in a channel},
    url = {https://www.sciencedirect.com/science/article/pii/S0309170803001593},
    doi = {10.1016/j.advwatres.2003.10.002},
    number = {1},
    journal = {Advances in Water Resources},
    author = {Kang, Qinjun and Zhang, Dongxiao and Chen, Shiyi},
    month = jan,
    year = {2004},
    pages = {13--22},
}

@article{lenormand_numerical_1988,
    title = {Numerical models and experiments on immiscible displacements in porous media},
    volume = {189},
    issn = {1469-7645, 0022-1120},
    doi = {10.1017/S0022112088000953},
    language = {en},
    journal = {Journal of Fluid Mechanics},
    author = {Lenormand, Roland and Touboul, Eric and Zarcone, Cesar},
    month = apr,
    year = {1988},
    pages = {165--187},
}

@article{mccracken_multiple-relaxation-time_2005,
    title = {Multiple-relaxation-time lattice-{Boltzmann} model for multiphase flow},
    volume = {71},
    url = {https://link.aps.org/doi/10.1103/PhysRevE.71.036701},
    doi = {10.1103/PhysRevE.71.036701},
    number = {3},
    journal = {Physical Review E},
    author = {McCracken, Michael E. and Abraham, John},
    month = mar,
    year = {2005},
    note = {Publisher: American Physical Society},
    pages = {036701},
}

@article{ataei_xlb_2024,
    title = {{XLB}: {A} differentiable massively parallel lattice {Boltzmann} library in {Python}},
    volume = {300},
    issn = {0010-4655},
    shorttitle = {{XLB}},
    url = {https://www.sciencedirect.com/science/article/pii/S0010465524001103},
    doi = {10.1016/j.cpc.2024.109187},
    journal = {Computer Physics Communications},
    author = {Ataei, Mohammadmehdi and Salehipour, Hesam},
    month = jul,
    year = {2024},
    pages = {109187},
}

@article{lou_evaluation_2013,
    title = {Evaluation of outflow boundary conditions for two-phase lattice {Boltzmann} equation},
    volume = {87},
    url = {https://link.aps.org/doi/10.1103/PhysRevE.87.063301},
    doi = {10.1103/PhysRevE.87.063301},
    number = {6},
    journal = {Physical Review E},
    author = {Lou, Qin and Guo, Zhaoli and Shi, Baochang},
    month = jun,
    year = {2013},
    note = {Publisher: American Physical Society},
    pages = {063301},
}

@misc{bedrunka_machine_2024,
    title = {Machine {Learning} {Enhanced} {Collision} {Operator} for the {Lattice} {Boltzmann} {Method} {Based} on {Invariant} {Networks}},
    url = {http://arxiv.org/abs/2412.08229},
    doi = {10.48550/arXiv.2412.08229},
    publisher = {arXiv},
    author = {Bedrunka, Mario Christopher and Horstmann, Tobias and Picard, Ben and Reith, Dirk and Foysi, Holger},
    month = dec,
    year = {2024},
    note = {arXiv:2412.08229 [physics]},
}

@article{horstmann_lattice_2024,
    title = {Lattice {Boltzmann} method with artificial bulk viscosity using a neural collision operator},
    volume = {272},
    issn = {0045-7930},
    url = {https://www.sciencedirect.com/science/article/pii/S0045793024000239},
    doi = {10.1016/j.compfluid.2024.106191},
    journal = {Computers \& Fluids},
    author = {Horstmann, Jan Tobias and Bedrunka, Mario Christopher and Foysi, Holger},
    month = mar,
    year = {2024},
    pages = {106191},
}

@article{wang_ml-lbm_2021,
    title = {{ML}-{LBM}: {Predicting} and {Accelerating} {Steady} {State} {Flow} {Simulation} in {Porous} {Media} with {Convolutional} {Neural} {Networks}},
    volume = {138},
    issn = {0169-3913, 1573-1634},
    shorttitle = {{ML}-{LBM}},
    url = {https://link.springer.com/10.1007/s11242-021-01590-6},
    doi = {10.1007/s11242-021-01590-6},
    language = {en},
    number = {1},
    journal = {Transport in Porous Media},
    author = {Wang, Ying Da and Chung, Traiwit and Armstrong, Ryan T. and Mostaghimi, Peyman},
    month = may,
    year = {2021},
    pages = {49--75},
}

@article{chaaban_machine-learning_2024,
    title = {A machine-learning supported multi-scale {LBM}-{TPM} model of unsaturated, anisotropic, and deformable porous materials},
    volume = {48},
    issn = {1096-9853},
    url = {https://onlinelibrary.wiley.com/doi/abs/10.1002/nag.3668},
    doi = {10.1002/nag.3668},
    language = {en},
    number = {4},
    journal = {International Journal for Numerical and Analytical Methods in Geomechanics},
    author = {Chaaban, Mohamad and Heider, Yousef and Sun, WaiChing and Markert, Bernd},
    year = {2024},
    pages = {889--910},
}

@article{graczyk_predicting_2020,
    title = {Predicting porosity, permeability, and tortuosity of porous media from images by deep learning},
    volume = {10},
    copyright = {2020 The Author(s)},
    issn = {2045-2322},
    url = {https://www.nature.com/articles/s41598-020-78415-x},
    doi = {10.1038/s41598-020-78415-x},
    language = {en},
    number = {1},
    journal = {Scientific Reports},
    author = {Graczyk, Krzysztof M. and Matyka, Maciej},
    month = dec,
    year = {2020},
    note = {Publisher: Nature Publishing Group},
    pages = {21488},
}

@article{fei_pore-scale_2022,
    title = {Pore-{Scale} {Study} on {Convective} {Drying} of {Porous} {Media}},
    volume = {38},
    issn = {0743-7463},
    url = {https://doi.org/10.1021/acs.langmuir.2c00267},
    doi = {10.1021/acs.langmuir.2c00267},
    number = {19},
    journal = {Langmuir},
    author = {Fei, Linlin and Qin, Feifei and Zhao, Jianlin and Derome, Dominique and Carmeliet, Jan},
    month = may,
    year = {2022},
    note = {Publisher: American Chemical Society},
    pages = {6023--6035},
}

@article{luo_unified_2021,
    title = {A unified lattice {Boltzmann} model and application to multiphase flows},
    volume = {379},
    issn = {1364-503X, 1471-2962},
    url = {https://royalsocietypublishing.org/doi/10.1098/rsta.2020.0397},
    doi = {10.1098/rsta.2020.0397},
    language = {en},
    number = {2208},
    journal = {Philosophical Transactions of the Royal Society A: Mathematical, Physical and Engineering Sciences},
    author = {Luo, Kai H. and Fei, Linlin and Wang, Geng},
    month = oct,
    year = {2021},
    pages = {20200397},
}

@article{tian_lattice_2025,
    title = {Lattice {Boltzmann} modeling of evaporation of porous media considering conjugate heat transfer},
    volume = {37},
    issn = {1070-6631, 1089-7666},
    url = {https://pubs.aip.org/pof/article/37/3/033351/3339912/Lattice-Boltzmann-modeling-of-evaporation-of},
    doi = {10.1063/5.0261217},
    language = {en},
    number = {3},
    journal = {Physics of Fluids},
    author = {Tian, Zhixing and Fei, Linlin and Wang, Chenglong and Guo, Kailun and Tian, Wenxi and Qiu, Suizheng and Su, Guanghui and Derome, Dominique and Carmeliet, Jan},
    month = mar,
    year = {2025},
    pages = {033351},
}

@article{song_lattice_2016,
    title = {A lattice {Boltzmann} model for heat and mass transfer phenomena with phase transformations in unsaturated soil during freezing process},
    volume = {94},
    issn = {0017-9310},
    url = {https://www.sciencedirect.com/science/article/pii/S0017931015010455},
    doi = {10.1016/j.ijheatmasstransfer.2015.11.008},
    journal = {International Journal of Heat and Mass Transfer},
    author = {Song, Wenyu and Zhang, Yaning and Li, Bingxi and Fan, Xinmeng},
    month = mar,
    year = {2016},
    pages = {29--38},
}

@article{zhou_pore-scale_2021,
    title = {Pore-{Scale} {Simulations} of {Particles} {Migration} and {Deposition} in {Porous} {Media} {Using} {LBM}-{DEM} {Coupling} {Method}},
    volume = {9},
    copyright = {http://creativecommons.org/licenses/by/3.0/},
    issn = {2227-9717},
    url = {https://www.mdpi.com/2227-9717/9/3/465},
    doi = {10.3390/pr9030465},
    language = {en},
    number = {3},
    journal = {Processes},
    author = {Zhou, Yanjie and Chen, Liping and Gong, Yanfeng and Wang, Shilin},
    month = mar,
    year = {2021},
    note = {Number: 3
Publisher: Multidisciplinary Digital Publishing Institute},
    pages = {465},
}

@article{gao_reactive_2017,
    title = {Reactive transport in porous media for {CO2} sequestration: {Pore} scale modeling using the lattice {Boltzmann} method},
    volume = {98},
    issn = {0098-3004},
    shorttitle = {Reactive transport in porous media for {CO2} sequestration},
    url = {https://www.sciencedirect.com/science/article/pii/S0098300416304307},
    doi = {10.1016/j.cageo.2016.09.008},
    journal = {Computers \& Geosciences},
    author = {Gao, Jinfang and Xing, Huilin and Tian, Zhiwei and Pearce, Julie K. and Sedek, Mohamed and Golding, Suzanne D. and Rudolph, Victor},
    month = jan,
    year = {2017},
    pages = {9--20},
}

@article{yoshino_lattice_2022,
    title = {{LATTICE} {BOLTZMANN} {SIMULATION} {OF} {BEHAVIORS} {OF} {BINARY} {CLOUD} {DROPLETS} {APPROACHING} {EACH} {OTHER}},
    volume = {34},
    issn = {0276-1459, 1943-6181},
    url = {https://www.dl.begellhouse.com/journals/5af8c23d50e0a883,2e84b81900eabcea,3de76a7e0e538f5d.html},
    doi = {10.1615/MultScienTechn.2022043604},
    language = {English},
    number = {3},
    journal = {Multiphase Science and Technology},
    author = {Yoshino, Masato and Sasaki, Kohei and Saito, Satoshi and Suzuki, Kosuke},
    year = {2022},
    note = {Publisher: Begel House Inc.},
}

@article{linhao_wind-driven_2005,
    title = {Wind-driven ocean circulation in shallow water lattice {Boltzmann} model},
    volume = {22},
    issn = {1861-9533},
    url = {https://doi.org/10.1007/BF02918749},
    doi = {10.1007/BF02918749},
    language = {en},
    number = {3},
    journal = {Advances in Atmospheric Sciences},
    author = {Linhao, Zhong and Shide, Feng and Shouting, Gao},
    month = jun,
    year = {2005},
    pages = {349--358},
}

@article{galindo-torres_boundary_2016,
    title = {Boundary effects on the {Soil} {Water} {Characteristic} {Curves} obtained from lattice {Boltzmann} simulations},
    volume = {71},
    issn = {0266-352X},
    url = {https://www.sciencedirect.com/science/article/pii/S0266352X15002062},
    doi = {10.1016/j.compgeo.2015.09.008},
    journal = {Computers and Geotechnics},
    author = {Galindo-Torres, S. A. and Scheuermann, A. and Li, L.},
    month = jan,
    year = {2016},
    pages = {136--146},
}

@article{porter_lattice-boltzmann_2009,
    title = {Lattice-{Boltzmann} simulations of the capillary pressure–saturation–interfacial area relationship for porous media},
    volume = {32},
    copyright = {https://www.elsevier.com/tdm/userlicense/1.0/},
    issn = {03091708},
    url = {https://linkinghub.elsevier.com/retrieve/pii/S0309170809001328},
    doi = {10.1016/j.advwatres.2009.08.009},
    language = {en},
    number = {11},
    journal = {Advances in Water Resources},
    author = {Porter, Mark L. and Schaap, Marcel G. and Wildenschild, Dorthe},
    month = nov,
    year = {2009},
    pages = {1632--1640},
}

@article{bezgin_jax-fluids_2025,
    title = {{JAX}-{Fluids} 2.0: {Towards} {HPC} for differentiable {CFD} of compressible two-phase flows},
    volume = {308},
    issn = {0010-4655},
    shorttitle = {{JAX}-{Fluids} 2.0},
    url = {https://www.sciencedirect.com/science/article/pii/S0010465524003564},
    doi = {10.1016/j.cpc.2024.109433},
    journal = {Computer Physics Communications},
    author = {Bezgin, Deniz A. and Buhendwa, Aaron B. and Adams, Nikolaus A.},
    month = mar,
    year = {2025},
    pages = {109433},
}

@article{premnath_dynamic_2009,
    title = {Dynamic subgrid scale modeling of turbulent flows using lattice-{Boltzmann} method},
    volume = {388},
    issn = {0378-4371},
    url = {https://www.sciencedirect.com/science/article/pii/S0378437109001745},
    doi = {10.1016/j.physa.2009.02.041},
    number = {13},
    journal = {Physica A: Statistical Mechanics and its Applications},
    author = {Premnath, Kannan N. and Pattison, Martin J. and Banerjee, Sanjoy},
    month = jul,
    year = {2009},
    pages = {2640--2658},
}

@article{bartlett_lattice_2013,
    title = {Lattice {Boltzmann} two-equation model for turbulence simulations: {High}-{Reynolds} number flow past circular cylinder},
    volume = {42},
    issn = {0142-727X},
    shorttitle = {Lattice {Boltzmann} two-equation model for turbulence simulations},
    url = {https://www.sciencedirect.com/science/article/pii/S0142727X13000325},
    doi = {10.1016/j.ijheatfluidflow.2013.01.018},
    journal = {International Journal of Heat and Fluid Flow},
    author = {Bartlett, Casey and Chen, Hudong and Staroselsky, Ilya and Wanderer, John and Yakhot, Victor},
    month = aug,
    year = {2013},
    pages = {1--9},
}

@article{filippova_multiscale_2001,
    title = {Multiscale {Lattice} {Boltzmann} {Schemes} with {Turbulence} {Modeling}},
    volume = {170},
    issn = {0021-9991},
    url = {https://www.sciencedirect.com/science/article/pii/S0021999101967646},
    doi = {10.1006/jcph.2001.6764},
    number = {2},
    journal = {Journal of Computational Physics},
    author = {Filippova, Olga and Succi, Sauro and Mazzocco, Francesco and Arrighetti, Cinzio and Bella, Gino and Hänel, Dieter},
    month = jul,
    year = {2001},
    pages = {812--829},
}

@article{li_non-body-fitted_2012,
    title = {Non-body-fitted {Cartesian}-mesh simulation of highly turbulent flows using multi-relaxation-time lattice {Boltzmann} method},
    volume = {63},
    issn = {0898-1221},
    url = {https://www.sciencedirect.com/science/article/pii/S0898122112002854},
    doi = {10.1016/j.camwa.2012.03.080},
    number = {10},
    journal = {Computers \& Mathematics with Applications},
    author = {Li, Kai and Zhong, Chengwen and Zhuo, Congshan and Cao, Jun},
    month = may,
    year = {2012},
    pages = {1481--1496},
}

@article{pellerin_implementation_2015,
    title = {An implementation of the {Spalart}–{Allmaras} turbulence model in a multi-domain lattice {Boltzmann} method for solving turbulent airfoil flows},
    volume = {70},
    issn = {0898-1221},
    url = {https://www.sciencedirect.com/science/article/pii/S0898122115004897},
    doi = {10.1016/j.camwa.2015.10.006},
    number = {12},
    journal = {Computers \& Mathematics with Applications},
    author = {Pellerin, Nicolas and Leclaire, Sébastien and Reggio, Marcelo},
    month = dec,
    year = {2015},
    pages = {3001--3018},
}

@article{karlin_gibbs_2014,
    title = {Gibbs' principle for the lattice-kinetic theory of fluid dynamics},
    volume = {90},
    url = {https://link.aps.org/doi/10.1103/PhysRevE.90.031302},
    doi = {10.1103/PhysRevE.90.031302},
    number = {3},
    journal = {Physical Review E},
    author = {Karlin, I. V. and Bösch, F. and Chikatamarla, S. S.},
    month = sep,
    year = {2014},
    note = {Publisher: American Physical Society},
    pages = {031302},
}

@article{kochkov_machine_2021,
    title = {Machine learning–accelerated computational fluid dynamics},
    volume = {118},
    url = {https://www.pnas.org/doi/abs/10.1073/pnas.2101784118},
    doi = {10.1073/pnas.2101784118},
    number = {21},
    journal = {Proceedings of the National Academy of Sciences},
    author = {Kochkov, Dmitrii and Smith, Jamie A. and Alieva, Ayya and Wang, Qing and Brenner, Michael P. and Hoyer, Stephan},
    month = may,
    year = {2021},
    note = {Publisher: Proceedings of the National Academy of Sciences},
    pages = {e2101784118},
}

@article{vinuesa_enhancing_2022,
    title = {Enhancing computational fluid dynamics with machine learning},
    volume = {2},
    copyright = {2022 Springer Nature America, Inc.},
    issn = {2662-8457},
    url = {https://www.nature.com/articles/s43588-022-00264-7},
    doi = {10.1038/s43588-022-00264-7},
    language = {en},
    number = {6},
    journal = {Nature Computational Science},
    author = {Vinuesa, Ricardo and Brunton, Steven L.},
    month = jun,
    year = {2022},
    note = {Publisher: Nature Publishing Group},
    pages = {358--366},
}

@article{coveney_multiplerelaxationtime_2002,
    title = {Multiple–relaxation–time lattice {Boltzmann} models in three dimensions},
    volume = {360},
    url = {https://royalsocietypublishing.org/doi/10.1098/rsta.2001.0955},
    doi = {10.1098/rsta.2001.0955},
    number = {1792},
    journal = {Philosophical Transactions of the Royal Society of London. Series A: Mathematical, Physical and Engineering Sciences},
    author = {Coveney, P. V. and Succi, S. and d'Humières, Dominique and Ginzburg, Irina and Krafczyk, Manfred and Lallemand, Pierre and Luo, Li-Shi},
    month = mar,
    year = {2002},
    note = {Publisher: Royal Society},
    pages = {437--451},
}

@article{geier_cascaded_2006,
    title = {Cascaded digital lattice {Boltzmann} automata for high {Reynolds} number flow},
    volume = {73},
    url = {https://link.aps.org/doi/10.1103/PhysRevE.73.066705},
    doi = {10.1103/PhysRevE.73.066705},
    number = {6},
    journal = {Physical Review E},
    author = {Geier, Martin and Greiner, Andreas and Korvink, Jan G.},
    month = jun,
    year = {2006},
    note = {Publisher: American Physical Society},
    pages = {066705},
}

@article{fei_consistent_2017,
    title = {Consistent forcing scheme in the cascaded lattice {Boltzmann} method},
    volume = {96},
    copyright = {https://link.aps.org/licenses/aps-default-license},
    issn = {2470-0045, 2470-0053},
    url = {https://link.aps.org/doi/10.1103/PhysRevE.96.053307},
    doi = {10.1103/PhysRevE.96.053307},
    language = {en},
    number = {5},
    journal = {Physical Review E},
    author = {Fei, Linlin and Luo, Kai Hong},
    month = nov,
    year = {2017},
    pages = {053307},
}

@article{ashirbekov_equation_2021,
    title = {Equation of {State}’s {Crossover} {Enhancement} of {Pseudopotential} {Lattice} {Boltzmann} {Modeling} of {CO2} {Flow} in {Homogeneous} {Porous} {Media}},
    volume = {6},
    copyright = {© 2021 by the authors.  Licensee MDPI, Basel, Switzerland. This article is an open access article distributed under the terms and conditions of the Creative Commons Attribution (CC BY) license (https://creativecommons.org/licenses/by/4.0/).  Notwithstanding the ProQuest Terms and Conditions, you may use this content in accordance with the terms of the License.},
    url = {https://www.proquest.com/docview/2612768462/abstract/E2EE3708F3804330PQ/1},
    doi = {10.3390/fluids6120434},
    language = {English},
    number = {12},
    journal = {Fluids},
    author = {Ashirbekov, Assetbek and Kabdenova, Bagdagul and Monaco, Ernesto and Rojas-Solórzano, Luis R.},
    year = {2021},
    note = {Num Pages: 434
Place: Basel, Switzerland
Publisher: MDPI AG},
    pages = {434},
}

@article{gunstensen_lattice_1991,
    title = {Lattice {Boltzmann} model of immiscible fluids},
    volume = {43},
    url = {https://link.aps.org/doi/10.1103/PhysRevA.43.4320},
    doi = {10.1103/PhysRevA.43.4320},
    number = {8},
    journal = {Physical Review A},
    author = {Gunstensen, Andrew K. and Rothman, Daniel H. and Zaleski, Stéphane and Zanetti, Gianluigi},
    month = apr,
    year = {1991},
    note = {Publisher: American Physical Society},
    pages = {4320--4327},
}

@article{swift_lattice_1996,
    title = {Lattice {Boltzmann} simulations of liquid-gas and binary fluid systems},
    volume = {54},
    url = {https://link.aps.org/doi/10.1103/PhysRevE.54.5041},
    doi = {10.1103/PhysRevE.54.5041},
    number = {5},
    journal = {Physical Review E},
    author = {Swift, Michael R. and Orlandini, E. and Osborn, W. R. and Yeomans, J. M.},
    month = nov,
    year = {1996},
    note = {Publisher: American Physical Society},
    pages = {5041--5052},
}

@article{xie_relative_2017,
    title = {Relative permeabilities of supercritical {CO2} and brine in carbon sequestration by a two-phase lattice {Boltzmann} method},
    volume = {53},
    issn = {0947-7411, 1432-1181},
    url = {http://link.springer.com/10.1007/s00231-017-2007-6},
    doi = {10.1007/s00231-017-2007-6},
    language = {en},
    number = {8},
    journal = {Heat and Mass Transfer},
    author = {Xie, Jian.-Fei. and He, S. and Zu, Y. Q. and Lamy-Chappuis, B. and Yardley, B. W. D.},
    month = aug,
    year = {2017},
    pages = {2637--2649},
}

@article{an_lattice-boltzmann_2021,
    title = {Lattice-{Boltzmann} simulation of dissolution of carbonate rock during {CO2}-saturated brine injection},
    volume = {408},
    issn = {1385-8947},
    url = {https://www.sciencedirect.com/science/article/pii/S1385894720333581},
    doi = {10.1016/j.cej.2020.127235},
    journal = {Chemical Engineering Journal},
    author = {An, Senyou and Erfani, Hamidreza and Hellevang, Helge and Niasar, Vahid},
    month = mar,
    year = {2021},
    pages = {127235},
}

@article{liu_review_2021,
    title = {A {Review} of {Lattice}-{Boltzmann} {Models} {Coupled} with {Geochemical} {Modeling} {Applied} for {Simulation} of {Advanced} {Waterflooding} and {Enhanced} {Oil} {Recovery} {Processes}},
    volume = {35},
    copyright = {https://doi.org/10.15223/policy-029},
    issn = {0887-0624, 1520-5029},
    url = {https://pubs.acs.org/doi/10.1021/acs.energyfuels.1c01347},
    doi = {10.1021/acs.energyfuels.1c01347},
    language = {en},
    number = {17},
    journal = {Energy \& Fuels},
    author = {Liu, Siyan and Zhang, Chi and Ghahfarokhi, Reza Barati},
    month = sep,
    year = {2021},
    pages = {13535--13549},
}

@article{zhang_flow_2019,
    title = {Flow {Mechanism} and {Simulation} {Approaches} for {Shale} {Gas} {Reservoirs}: {A} {Review}},
    volume = {126},
    issn = {0169-3913, 1573-1634},
    shorttitle = {Flow {Mechanism} and {Simulation} {Approaches} for {Shale} {Gas} {Reservoirs}},
    url = {http://link.springer.com/10.1007/s11242-018-1148-5},
    doi = {10.1007/s11242-018-1148-5},
    language = {en},
    number = {3},
    journal = {Transport in Porous Media},
    author = {Zhang, Tao and Sun, Shuyu and Song, Hongqing},
    month = feb,
    year = {2019},
    pages = {655--681},
}

@article{yang_characterizing_2025,
    title = {Characterizing {Dynamic} {Contact} {Angle} during {Gas}–{Liquid} {Imbibition} in {Microchannels} by {Lattice} {Boltzmann} {Method} {Modeling}},
    volume = {10},
    copyright = {https://creativecommons.org/licenses/by-nc-nd/4.0/},
    issn = {2470-1343, 2470-1343},
    url = {https://pubs.acs.org/doi/10.1021/acsomega.4c10365},
    doi = {10.1021/acsomega.4c10365},
    language = {en},
    number = {3},
    journal = {ACS Omega},
    author = {Yang, Xuefeng and Chang, Cheng and Zheng, Majia and Wang, Xingchen and Chen, Yizhao and Xie, Weiyang and Hu, Haoran and Cheng, Qiuyang},
    month = jan,
    year = {2025},
    pages = {3116--3127},
}

@article{wolf_capillary_2010,
    title = {Capillary rise between parallel plates under dynamic conditions},
    volume = {344},
    copyright = {https://www.elsevier.com/tdm/userlicense/1.0/},
    issn = {00219797},
    url = {https://linkinghub.elsevier.com/retrieve/pii/S0021979709015604},
    doi = {10.1016/j.jcis.2009.12.023},
    language = {en},
    number = {1},
    journal = {Journal of Colloid and Interface Science},
    author = {Wolf, Fabiano G. and Dos Santos, Luís O.E. and Philippi, Paulo C.},
    month = apr,
    year = {2010},
    pages = {171--179},
}

@article{washburn_dynamics_1921,
    title = {The {Dynamics} of {Capillary} {Flow}},
    volume = {17},
    url = {https://link.aps.org/doi/10.1103/PhysRev.17.273},
    doi = {10.1103/PhysRev.17.273},
    number = {3},
    journal = {Physical Review},
    author = {Washburn, Edward W.},
    month = mar,
    year = {1921},
    note = {Publisher: American Physical Society},
    pages = {273--283},
}

@article{hamraoui_can_2000,
    title = {Can a {Dynamic} {Contact} {Angle} {Be} {Understood} in {Terms} of a {Friction} {Coefficient}?},
    volume = {226},
    copyright = {https://www.elsevier.com/tdm/userlicense/1.0/},
    issn = {00219797},
    url = {https://linkinghub.elsevier.com/retrieve/pii/S0021979700968309},
    doi = {10.1006/jcis.2000.6830},
    language = {en},
    number = {2},
    journal = {Journal of Colloid and Interface Science},
    author = {Hamraoui, Ahmed and Thuresson, Krister and Nylander, Tommy and Yaminsky, Vassili},
    month = jun,
    year = {2000},
    pages = {199--204},
}

@article{siebold_effect_2000,
    title = {Effect of dynamic contact angle on capillary rise phenomena},
    volume = {161},
    copyright = {https://www.elsevier.com/tdm/userlicense/1.0/},
    issn = {09277757},
    url = {https://linkinghub.elsevier.com/retrieve/pii/S0927775799003271},
    doi = {10.1016/S0927-7757(99)00327-1},
    language = {en},
    number = {1},
    journal = {Colloids and Surfaces A: Physicochemical and Engineering Aspects},
    author = {Siebold, Alain and Nardin, Michel and Schultz, Jacques and Walliser, André and Oppliger, Max},
    month = jan,
    year = {2000},
    pages = {81--87},
}

@article{heshmati_experimental_2014,
    title = {Experimental {Investigation} of {Dynamic} {Contact} {Angle} and {Capillary} {Rise} in {Tubes} with {Circular} and {Noncircular} {Cross} {Sections}},
    volume = {30},
    issn = {0743-7463, 1520-5827},
    url = {https://pubs.acs.org/doi/10.1021/la501724y},
    doi = {10.1021/la501724y},
    language = {en},
    number = {47},
    journal = {Langmuir},
    author = {Heshmati, Mohammad and Piri, Mohammad},
    month = dec,
    year = {2014},
    pages = {14151--14162},
}

@book{sukop_lattice_2006,
    address = {Berlin, Heidelberg},
    title = {Lattice {Boltzmann} {Modeling}: {An} {Introduction} for {Geoscientists} and {Engineers}},
    isbn = {978-3-540-27981-5},
    shorttitle = {Lattice {Boltzmann} {Modeling}},
    url = {http://link.springer.com/10.1007/978-3-540-27982-2},
    language = {en},
    publisher = {Springer},
    author = {Sukop, Michael C. and Thorne, Daniel T.},
    year = {2006},
    doi = {10.1007/978-3-540-27982-2},
}

@article{xu_pore-scale_2020,
    title = {Pore-{Scale} {Study} of {Water} {Adsorption} and {Subsequent} {Methane} {Transport} in {Clay} in the {Presence} of {Wettability} {Heterogeneity}},
    volume = {56},
    copyright = {©2020. American Geophysical Union. All Rights Reserved.},
    issn = {1944-7973},
    url = {https://onlinelibrary.wiley.com/doi/abs/10.1029/2020WR027568},
    doi = {10.1029/2020WR027568},
    language = {en},
    number = {10},
    journal = {Water Resources Research},
    author = {Xu, Rui and Prodanović, Maša and Landry, Christopher},
    year = {2020},
    pages = {e2020WR027568},
}

@book{wang_machine_2024,
    title = {Machine {Learning} {Assisting} {Intelligent} {Control} of {Evaporation} {Performance} in {Porous} {Media}},
    isbn = {978-0-85014-726-1},
    url = {https://www.intechopen.com/online-first/1194321},
    language = {en},
    publisher = {IntechOpen},
    author = {Wang, Hui and Xu, Shaoxuan and Qu, Zhiguo and Wang, Hui and Xu, Shaoxuan and Qu, Zhiguo},
    month = oct,
    year = {2024},
    doi = {10.5772/intechopen.1007473},
}

@article{pitsch_large-eddy_2006,
    title = {{LARGE}-{EDDY} {SIMULATION} {OF} {TURBULENT} {COMBUSTION}},
    volume = {38},
    issn = {0066-4189, 1545-4479},
    url = {https://www.annualreviews.org/content/journals/10.1146/annurev.fluid.38.050304.092133},
    doi = {10.1146/annurev.fluid.38.050304.092133},
    language = {en},
    number = {Volume 38, 2006},
    journal = {Annual Review of Fluid Mechanics},
    author = {Pitsch, Heinz},
    month = jan,
    year = {2006},
    note = {Publisher: Annual Reviews},
    pages = {453--482},
}

@article{kelly_assessing_2016,
    title = {Assessing the utility of {FIB}-{SEM} images for shale digital rock physics},
    volume = {95},
    issn = {03091708},
    url = {https://linkinghub.elsevier.com/retrieve/pii/S0309170815001360},
    doi = {10.1016/j.advwatres.2015.06.010},
    language = {en},
    journal = {Advances in Water Resources},
    author = {Kelly, Shaina and El-Sobky, Hesham and Torres-Verdín, Carlos and Balhoff, Matthew T.},
    month = sep,
    year = {2016},
    pages = {302--316},
}

@article{yang_recent_2023,
    title = {Recent advances on fluid flow in porous media using digital core analysis technology},
    volume = {9},
    issn = {2207-9963},
    url = {https://www.sciopen.com/article/10.46690/ager.2023.08.01},
    doi = {10.46690/ager.2023.08.01},
    language = {en},
    number = {2},
    urldate = {2025-06-17},
    journal = {Advances in Geo-Energy Research},
    author = {Yang, Yongfei and Horne, Roland N. and Cai, Jianchao and Yao, Jun},
    month = aug,
    year = {2023},
    pages = {71--75},
}

@misc{kidger_equinox_2021,
    title = {Equinox: neural networks in {JAX} via callable {PyTrees} and filtered transformations},
    shorttitle = {Equinox},
    url = {http://arxiv.org/abs/2111.00254},
    doi = {10.48550/arXiv.2111.00254},
    urldate = {2025-10-23},
    publisher = {arXiv},
    author = {Kidger, Patrick and Garcia, Cristian},
    month = oct,
    year = {2021},
    note = {arXiv:2111.00254 [cs]},
}

@misc{kingma_adam_2017,
    title = {Adam: {A} {Method} for {Stochastic} {Optimization}},
    shorttitle = {Adam},
    url = {http://arxiv.org/abs/1412.6980},
    doi = {10.48550/arXiv.1412.6980},
    urldate = {2025-10-23},
    publisher = {arXiv},
    author = {Kingma, Diederik P. and Ba, Jimmy},
    month = jan,
    year = {2017},
    note = {arXiv:1412.6980 [cs]},
}

@article{corbetta_toward_2023,
    title = {Toward learning {Lattice} {Boltzmann} collision operators},
    volume = {46},
    issn = {1292-8941, 1292-895X},
    url = {https://link.springer.com/10.1140/epje/s10189-023-00267-w},
    doi = {10.1140/epje/s10189-023-00267-w},
    language = {en},
    number = {3},
    urldate = {2025-04-05},
    journal = {The European Physical Journal E},
    author = {Corbetta, Alessandro and Gabbana, Alessandro and Gyrya, Vitaliy and Livescu, Daniel and Prins, Joost and Toschi, Federico},
    month = mar,
    year = {2023},
    pages = {10},
}

@article{kong_adjoint_2025,
    title = {Adjoint lattice {Boltzmann} method-based multiscale topology optimization of thermo-fluid systems using anisotropic porous media},
    volume = {37},
    issn = {1070-6631},
    url = {https://doi.org/10.1063/5.0279717},
    doi = {10.1063/5.0279717},
    number = {8},
    urldate = {2025-10-25},
    journal = {Physics of Fluids},
    author = {Kong, Xiangzhuang and Zhang, Hongming and Du, Yanxia and Wang, Xian and Xiao, Guangming},
    month = aug,
    year = {2025},
    pages = {087208},
}
\end{document}